\documentclass[12pt,preprint]{aastex}

\usepackage{natbib}
\usepackage{color}

\newcommand\MSUN{\rm M_{\odot}}

\newcommand\Trat{{T_{\rm i}/T_{\rm e}}}

\newcommand\GHz{{\rm\,GHz}}

\newcommand\Rg{{\,G M/c^2}}
\newcommand\sgra{{Sgr~A*}}

\newcommand\munit{{\mathcal M}}
\newcommand\lunit{{\mathcal L}}
\newcommand\tunit{{\mathcal T}}

\newcommand\ergps{{\rm ergs \, s^{-1}}}

\newcommand\cm{{\rm\,cm}}

\renewcommand\deg{{\rm\,deg}}

\newcommand\keV{{\rm\,keV}} 

\newcommand\gdet{{\rm\,\sqrt{-g}}} 
\newcommand\eps{{\rm\,\epsilon}}
\newcommand\ecm{{\rm\,\epsilon_{[{\rm CM}]}}} 

\newcommand\mpair{{e^{\pm} }}
\newcommand\pprod{$\dot{n}_{\pm}   $}
\newcommand\mpprod{\dot{n}_{\pm} }
\newcommand\dm{\dot{m}}

\begin{document}

\title{Pair production in low luminosity galactic nuclei}

\author{Monika Mo{\'s}cibrodzka\altaffilmark{1}, Charles
  F. Gammie\altaffilmark{1,2}, Joshua C. Dolence\altaffilmark{2}, \\  Hotaka Shiokawa\altaffilmark{2}}

\affil{$^1$ Department of Physics, University of Illinois, 1110 West Green
  Street, Urbana, IL 61801}
\affil{$^2$ Astronomy Department, University of Illinois, 1002 West Green
  Street, Urbana, IL 61801}
\email{mmosc@illinois.edu}

\begin{abstract}

Electron-positron pairs may be produced near accreting black holes by a
variety of physical processes, and the resulting pair plasma may be
accelerated and collimated into a relativistic jet.  Here we use a
self-consistent dynamical and radiative model to investigate pair
production by $\gamma\gamma$ collisions in weakly radiative accretion
flows around a black hole of mass $M$ and accretion rate $\dot{M}$.  Our
flow model is drawn from general relativistic magnetohydrodynamic
simulations, and our radiation field is computed by a Monte Carlo
transport scheme assuming the electron distribution function is thermal.
We argue that the pair production rate scales as $r^{-6} M^{-1}
\dot{M}^{6}$.  We confirm this numerically and calibrate the scaling
relation.  This relation is self-consistent in a wedge in $M,
\dot{M}$ parameter space.  If $\dot{M}$ is too low the implied pair
density over the poles of the black hole is below the Goldreich-Julian
density and $\gamma\gamma$ pair production is relatively unimportant; if
$\dot{M}$ is too high the models are radiatively efficient.  We also
argue that for a power-law spectrum the pair production rate should
scale with the observables $L_X \equiv$ X-ray luminosity and $M$ as
$L_X^2 M^{-4}$. We confirm this numerically and argue that this relation
likely holds even for radiatively efficient flows.  The pair production
rates are sensitive to black hole spin and to the ion-electron
temperature ratio which are fixed in this exploratory calculation.  We
finish with a brief discussion of the implications for Sgr A* and M87.

\end{abstract}

\section{Introduction} \label{intro}

Models of zero-obliquity black hole accretion---in which the accretion
flow angular momentum is parallel to the black hole spin---typically
exhibit a low density ``funnel'' over the poles of the black hole.  The
funnel is empty because the funnel plasma is free to fall into the hole
or be ejected to large radius.  Magnetic fields do not prevent this: in
magnetohydrodynamic (MHD) simulations of radiatively inefficient
accretion flows (RIAFs) the funnel magnetic field typically runs in a
smooth spiral from the event horizon to large radius
\citep{devilliers:2003,mckinney:2004,komissarov:2005,hawley:2006,
beckwith:2008}.  Because the field lines do not leave the funnel there
is no way for the disk plasma to resupply the funnel plasma.  

What process, then, populates the funnel with plasma?  And what controls
the temperature (or distribution function) of the funnel plasma?  These
questions bear directly on two interesting problems in black hole jet
theory: are jets made of pairs or an electron-ion plasma?  And which is
more luminous: the base of the jet or the accretion flow?  The purpose
of this paper is to investigate these questions in the specific context
of hot, underluminous accretion flows where nearly {\it ab initio}
models are computationally feasible.

There are several pair creation processes that might populate the funnel
with plasma.  Plasma close to the event horizon in a RIAF is
relativistically hot, and thus can form electron-positron pairs $\mpair$
through particle-particle ($ee$, $ep$), particle-photon ($e\gamma$,
$p\gamma$), or photon-photon collisions ($\gamma\gamma$). The cross
section near the $\mpair$ energy threshold is largest for $\gamma
\gamma$ interactions, which have a cross section $\sim \sigma_{T}
\equiv$, the Thomson cross section. In the funnel the photon density
vastly exceeds the particle density, so $\gamma\gamma$ collisions
dominate $\mpair$ production (\citealt{stepney:1983},
\citealt{phinney:1983}, \citealt{phinney:1995}, \citealt{krolik:1999}).
Pair production by these processes is discussed in e.g.
\citealt{kusunose:1996} and \citealt{esin:1999} in the context of
Advection Dominated Accretion Flows (ADAFs, \citealt{narayan:1994}).
These works, however, focus on the energetic role of pairs in ADAF disks
rather than the population and dynamics of pairs in the funnel.

Other processes are important when the density is below the
\cite{goldreich:1969} charge density.  Then the plasma can
have ${\mathbf E} \cdot {\mathbf B} \ne 0$, and the electric field can
directly accelerate particles to high Lorentz factors. The energetic
particles Compton upscatter background photons that collide with other
background photons and produce a shower of pairs in a pair-photon
cascade (\citealt{blandford:1977, phinney:1983, beskin:1992,
hirotani:1998}, and recently \citealt{vincent:2010}). 

In this paper we model production of an $\mpair$ plasma by photon-photon
collisions in the funnel above a hot, underluminous accretion disk.  At
low accretion rates $\dot{M}$ ($\lesssim \dot{M}_{crit} \sim 10^{-6}
L_{Edd}/(0.1 c^2)$, where $L_{Edd} \equiv$ Eddington luminosity) the
disk cools on a timescale longer than the accretion timescale; it is a
RIAF. In this regime the radiative and dynamical evolution are decoupled
and it is practical to treat both on a nearly {\it ab initio} basis.
Throughout the range of $\dot{M}$ we consider the funnel pair plasma is
tenuous enough that annhilation is negligible, so pair production will
be balanced by advective losses such as accretion into the black hole or
loss in a wind.

We draw our RIAF model from two and three dimensional general
relativistic magnetohydrodynamics simulations (GRMHD, using the HARM
code, \citealt{gammie:2003,noble:2009}) of an accreting, magnetized torus
with zero cooling. The radiation field is calculated as a
post-processing step using a Monte Carlo method ({\tt grmonty},
\citealt{dolence:2009}), and pair production rates are estimated from
snapshots of the radiation field using a procedure described in detail
below.

This paper is organized as follows. In \S~\ref{model_eq} we describe the
basic model for accretion flow dynamics and radiative transfer. In
\S~\ref{pair_eq} we write down the pair production model and present a
test problem for our Monte Carlo scheme.  Scaling formulas are presented
in \S~\ref{pair_scaling}.  In \S~\ref{results} we show results for
a range of black hole masses and accretion rates.  We briefly discuss
implications for \sgra\ and M87 in \S~\ref{discussion} and summarize in
\S~\ref{summary}.

\section{Accretion flow model}\label{model_eq}

We use a numerical model for the accretion flow and for the radiation
field; together these nearly {\em ab initio} models form a numerical
laboratory for investigating physical processes near a black hole in a
self-consistent way.  

\subsection{Dynamical model}

We use a relativistic MHD model for the accreting plasma \citep[see
e.g.,][]{gammie:2003}. The initial condition is an equilibrium torus
\citep{fishbone:1976} in orbit around a Kerr black hole with $a_* =
0.94$, where $a_* G M^2/c$ is the hole angular momentum. The torus is
seeded with poloidal, concentric loops of weak magnetic field that are
parallel to density contours.  Small perturbations are added to the
internal energy and this seeds the magnetorotational instability, which
leads to the development of MHD turbulence in the disk and accretion
onto the central black hole.  The model extends from slightly inside the
event horizon to $r = 40 GM/c^2$.

We solve the evolution equations until a quasi-equilibrium accretion
flow is established, meaning that the mean structure of the flow is not
evolving on the dynamical timescale.  Our (untested) hypothesis is that
at $r < 15 GM/c^2$ the model accurately represent the inner portions of
a relaxed accretion flow extending over many decades in radius.  

A few of the physical assumptions in the GRMHD model are worth stating
explicitly.  The equation of state is 
\begin{equation}
p = (\gamma_{ad} - 1) u
\end{equation}
where $\gamma_{ad} = 13/9$ (appropriate for ion temperature $T_i < m_p
c^2/k = 1.1 \times 10^{13} K$ and electron temperature $T_e > m_e c^2/k
= 5.9 \times 10^9 K$), $p \equiv$ pressure, and $u\equiv$ internal
energy density.  Particle number is also conserved:
\begin{equation}
(\rho_0 u^\mu)_{;\mu} = 0,
\end{equation}
where $\rho_0 \equiv$ rest-mass density and $u^\mu \equiv$
four-velocity, in the dynamical evolution. That is, pair production is
not included in the dynamical model.  The model is therefore consistent
only if pair creation is weak enough not to alter the flow dynamics or
energetics.  

\begin{figure*}
\begin{picture}(0,550)
\put(-140,550){\includegraphics{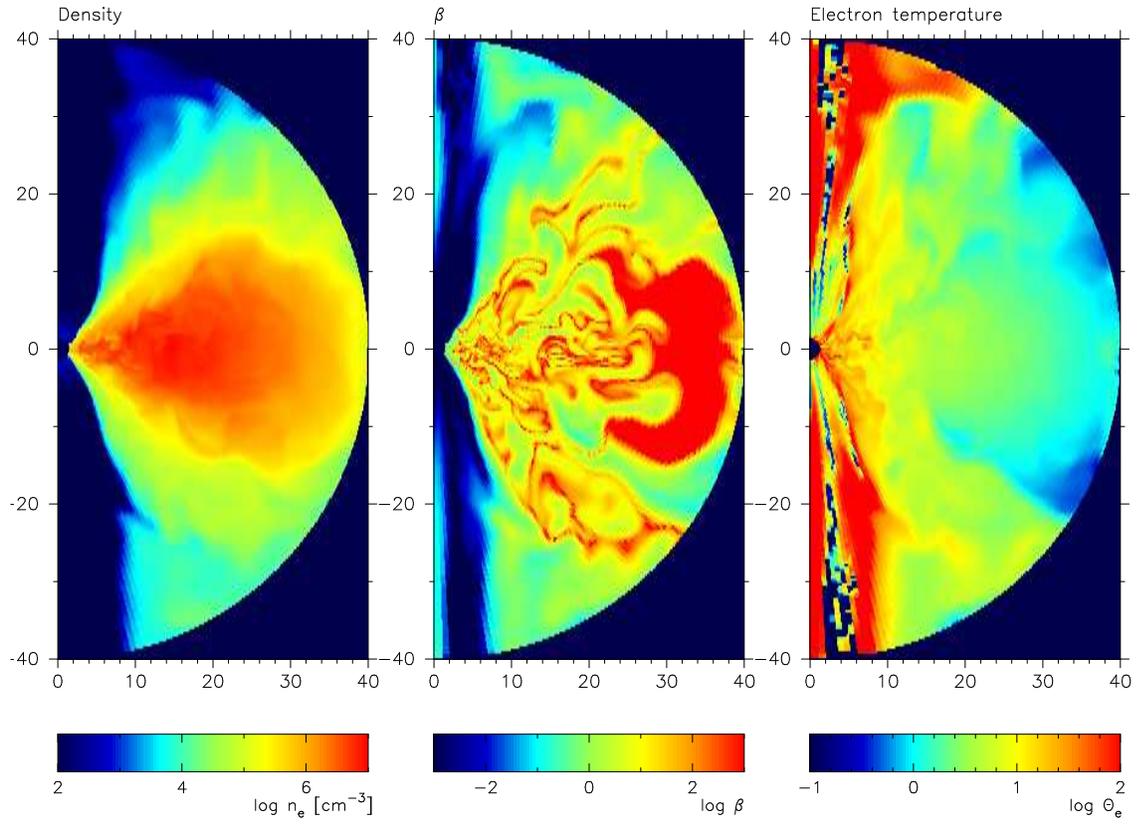}}
\end{picture}
\caption{\normalsize
Structure of RIAF. Panels from left to right:
density distribution, plasma $\beta$ parameter,
and dimensionless electron temperature, respectively.}
\label{fig:0}
\end{figure*}

We evolve the GRMHD equations using the {\tt harm} code
(\citealt{gammie:2003,noble:2009}). {\tt harm} is a conservative scheme that
evolves the total energy rather than internal energy of the flow. The
MHD equation integration is performed on a uniform grid in modified
Kerr-Schild coordinates \citep{gammie:2003}. The coordinates are
logarithmic in Kerr-Schild radius $r$ and nonuniform in Kerr-Schild
colatitude $\theta$ (Boyer-Lindquist and Kerr-Schild $r$ and $\theta$
are identical), with zones concentrated toward the midplane of the
accretion disk.  The inner and outer radial boundaries use outflow
boundary conditions.  The axisymmetric models use a $256 \times
256$ grid, and the single 3D run uses a $192 \times 192 \times 128$
grid.  For details of the numerical method, the initial setup and the
flow evolution in 2D see \citet{gammie:2003} and \citet{mckinney:2004}.
A snapshot of the density, temperature, and magnetic field strength from
one of our runs is shown in Figure~\ref{fig:0}.

\subsection{Radiative model}

Our radiative model is identical to that applied by
\citet{moscibrodzka:2009} to \sgra, although here we restrict attention
to a thermal plasma with $T_e = T_i$ (except in one case noted below).
Synchrotron emission and absorption are included, as is Compton
scattering.  

Bremsstrahlung is not important in the inner parts of the accretion
flow.  For a thermal plasma  with $\Theta_e \equiv k T_e/(m_e c^2) > 1$
the ratio (synchrotron / bremsstrahlung) cooling $\sim
\Theta_e^2/(\alpha\beta)$, where $\alpha \equiv$ fine structure constant
and $\beta \equiv 8\pi p/B^2$.  At the radii of interest here $\Theta_e
\sim 1 - 10^2$, and $\beta \sim 10$, so in an energetic sense
synchrotron dominates the direct production of photons.

Synchrotron emission occurs at a characteristic
frequency $\nu_s \sim (e B/(2 \pi m_e c)) \Theta_e^2$ which is $\ll m_e
c^2/h$ for any astrophysically reasonable combination of $M$ and
$\dot{M}$~\footnote{For the synchrotron emissivity we use the
approximate expression of \citet{leung:2010}
\begin{equation}
j_{\nu} = \frac{\sqrt{2}\pi e^2 n_e \nu_s}{3cK_2(\Theta_e^{-1})} (X^{1/2} +
2^{11/12} X^{1/6} )^2 \exp(-X^{1/3}) \label{emi_leung}
\end{equation}
where $X = \nu/\nu_s$, $\nu_s=2/9 (eB/2\pi m_ec) \Theta_e^2 \sin\theta$
is the synchrotron frequency, $\theta$ is an angle between the magnetic
field vector and emitted photon, and $K_2$ is a modified Bessel function
of the second kind.  The fractional error for this approximate formula
is smaller than 1\% for $\Theta_e \ge 1$ (where most of the emission
occurs) and increases to $10\%$ and more at low frequencies for
$\Theta_e \le 1$ (where there is very little emission).  The synchrotron
emissivity function peaks at $\nu \approx 8 \nu_s$.}.
Potentially pair-producing photons must therefore be produced by Compton
scattering.

We compute the radiation field using the general relativistic Monte Carlo
radiative transfer code {\tt grmonty} \citep{dolence:2009}.  The
radiation field is represented by photon packets (photon rays or
``superphotons'').  Each superphoton is characterized by a weight
$w$ = number of physical photons/superphoton, and a wave four-vector
$k^{\mu}$.  Superphotons are produced by sampling the emissivity.
The wavevector is transported according to the geodesic equation.
Along a geodesic $w$ is decremented to account for synchrotron
absorption.  Compton scattering is incorporated by sampling scattering
events. When a superphoton scatters it is divided into a scattered piece
with new wavevector $k'^\mu$ and new weight $w'$, and an unscattered
piece along the original wavevector with weight $w - w'$.  The
distribution of scattered $k'^\mu$ is consistent with the full
Klein-Nishina differential cross section.

We use a ``fast light'' approximation in treating the radiative
transfer.  The data from a single time slice $t_n$ (e.g. $\rho_0 (t_n,
x^1, x^2, x^3)$) is used to calculate the emergent radiation field as if
the data, and therefore photon field, were time-independent.  We have
checked the fast light model against a time-dependent radiative transfer
model \citep{dolence:2011} and verified that this approximation does not
introduce significant errors. 

\subsection{Model scaling}

The properties of the accretion flow model are independent of the
absolute value of the density (provided the magnetic field strength is
scaled appropriately), but the radiative model is not.  To scale
the model we specify the length unit
\begin{equation}
{\mathcal L} \equiv \frac{GM}{c^2},
\end{equation}
time unit
\begin{equation}
{\mathcal T} \equiv \frac{GM}{c^3},
\end{equation}
and mass unit ${\mathcal M}$, which is proportional to the mass
accretion rate. $M$ does not set a mass scale because it appears only in
the combination $G M$.  Since ${\mathcal M} \lll M$ the accretion flow
does not affect the gravitational field.  

Given $M$ and ${\mathcal M}$ the radiative transfer calculation is well
posed.  Typically $M$ can be estimated directly from observations, while
${\mathcal M}$ is varied until the model submillimeter flux matches the
observed flux.

\subsection{Model limitations}

An important limitation of our model is that it treats accreting plasma
as a nonradiating ideal fluid.  This implies that electrons and ions
have an isotropic, thermal distribution function.  The potentially
important effects of pressure anisotropy and conduction (e.g.
\citealt{sharma:2006}, \citealt{johnson:2007}) are therefore neglected,
as are the radiative effects of a nonthermal component in the electron
distribution function.

Cooling is also neglected. This is a good approximation in low accretion
rate systems like Sgr A*, but a poor approximation in higher accretion
rate systems like M87.  If one were to turn on cooling but hold the
synchrotron flux fixed the density and magnetic field strength (i.e. the
mass unit $\munit$) would increase.

\section{Pair production}\label{pair_eq}

\subsection{Basic equations}

For a population of photons with distribution function $dN_\gamma/d^3x d^3k$
(here $d^3k \equiv dk_1 dk_2 dk_3$ and $1,2,3$ are the spatial coordinates)
the invariant pair production rate per unit volume is
\begin{equation}
\dot{n}_\pm \equiv \frac{1}{\sqrt{-g}} \frac{dN_{\pm}}{d^3xdt} = \frac{1}{2}
\int \frac{d^3k}{\sqrt{-g} k^t} \frac{d^3k'}{\sqrt{-g} k'^t}
\frac{dN_\gamma}{d^3x d^3k} \frac{dN_\gamma}{d^3x d^3k'} \epsilon^2_{[CM]}
\sigma_{\gamma\gamma} c \label{eq:pair_creation_general}
\end{equation}
where $g$ is the determinant of $g_{\mu\nu}$, and the factor of $1/2$ prevents
double-counting.  Here $\sigma_{\gamma\gamma}$ is the cross section for
$\gamma+\gamma \rightarrow e^{+} + e^{-}$:
\begin{equation}
\frac{\sigma_{\gamma\gamma}}{\sigma_{T}}=\frac{3}{8 \ecm^6} \left[ (2 \ecm^4
  +2 \ecm^2-1) \cosh^{-1}\ecm - \ecm (\ecm^2+1)\sqrt{\ecm^2-1}\right]
\end{equation}
(Breit \& Wheeler 1936),
\begin{equation}
\ecm = -u_{CM \mu} k^{\mu} = -u_{[CM] \mu} k'^\mu = \left(\frac{-k_\mu k'^{\mu}}
     {2} \right)^{1/2} \label{eq:ecm}
\end{equation}
is the energy of either photon in the center of momentum ([CM]) frame of the
two photons, and $u_{[CM]}$ is the four-velocity of the [CM] frame.

Equation~\ref{eq:pair_creation_general} is coordinate invariant since
$\sqrt{-g}d^3x dt$ is invariant, the distribution function is invariant
(because $d^3x d^3k$ is invariant), $\ecm$ is a scalar, the cross
section is invariant, and $d^3k/\sqrt{-g} k^t$ is invariant.  It reduces
to the correct rate (cf. eq. 12.7 of Landau \& Lifshitz, Classical
Theory of Fields) in Minkowski space, and is therefore the correct
general expression for the pair production rate.  Because $\mpprod$
itself is invariant it also describes the pair creation rate in
the fluid frame.

We will need the rate of four-momentum transfer from the radiation
field to the plasma via pair creation:
\begin{equation}
G^\mu \equiv \frac{1}{\sqrt{-g}} \frac{dP^\mu_{\pm}}{d^3xdt} = A \frac{1}{2}
\int \frac{d^3k}{\sqrt{-g} k^t} \frac{d^3k'}{\sqrt{-g} k'^t}
\frac{dN_\gamma}{d^3x d^3k} \frac{dN_\gamma}{d^3x d^3k'} \left(k^\mu +
k'^\mu\right) \epsilon^2_{[CM]} \sigma_{\gamma\gamma} c.
\label{eq:pair_creation_momentum}
\end{equation}
Here $A$ is a constant that makes the equation dimensionally correct.

\subsection{Monte Carlo estimate of pair creation rate}

We estimate the integrals (\ref{eq:pair_creation_general}) and
(\ref{eq:pair_creation_momentum}) using a Monte Carlo scheme.  Given a
sample of photons on a time slice $t$ within a small three-volume
$\Delta^3 x$, a naive estimate is
\begin{equation}\label{NAIVEEQ}
\frac{1}{\sqrt{-g}} \frac{dN_{\pm}}{d^3x dt} \approx \frac{1}{2}
\sum_{i,j} \left(\frac{w_i}{\Delta^3x}\right) \left(\frac{w_j}{\Delta^3x}\right)
\frac{1}{\sqrt{-g} k^t_i} \frac{1}{\sqrt{-g} k^t_j} \epsilon^2_{[CM]}
\sigma_{\gamma\gamma} c.
\end{equation}
where $i$ and $j$ label superphotons.

If there are $N_s$ superphotons in $\Delta^3 x$ then there are
$O(N_s^2)$ pairs of superphotons and the computational cost of
(\ref{NAIVEEQ}) is $O(N_s^2)$.  One might hope that the error would
scale as $1/\sqrt{N_s^2}$ because there are $O(N_s^2)$ pairs, but this
is wrong.  There are only $N_s$ independent samples and so the error
scales as $1/\sqrt{N_s}$.  

We obtain an estimate with accuracy that is the same order as
(\ref{NAIVEEQ}) at $O(N_s)$ cost by selecting an unbiased sample of 
$N_s$ pairs of superphotons and evaluating
\begin{equation}
\frac{1}{\sqrt{-g}} \frac{dN_{\pm}}{d^3x dt} 
\approx 
\frac{1}{2} \frac{N_s}{2} 
\sum_{i,j} \left(\frac{w_i}{\Delta^3x}\right)
\left(\frac{w_j}{\Delta^3x}\right) \frac{1}{\sqrt{-g} k^t_i}
\frac{1}{\sqrt{-g} k^t_j} \epsilon^2_{[CM]} \sigma_{\gamma\gamma}
c \label{equation:pair}
\end{equation}
An identical procedure is used to evaluate $G^\mu$. 

\subsection{Test problem}

Does our Monte Carlo procedure accurately estimate the pair production
rate?  As a check we evaluate pair production rates near two point
sources of $E = 4 m_e c^2$ photons.  The calculation is done in
Minkowski space and Cartesian coordinates (so $\gdet = 1$), and the
optical depth to pair creation is assumed small.  At each point we
compare the analytic and numerical result.

The expected pair production rate is given by Equation~
\ref{eq:pair_creation_general}.  The energy of two colliding photons in
their center-of-momentum frame is a function of the cosine $\mu$ of the
angle between the rays from the two sources: $\epsilon_{[CM]}^2=(1/2) (1
- \mu) k^t k'^t$.  The photon momentum space distribution is a $\delta$
function.  The number density of photons $dN_{\gamma}/d^3x =
\dot{N}_\gamma/(4 \pi r^2 c)$ at distance $r$ from the source, where
each source produces photons at rate $\dot{N}_\gamma$.

\begin{figure*}
\begin{picture}(0,550)
\put(-240,50){\includegraphics{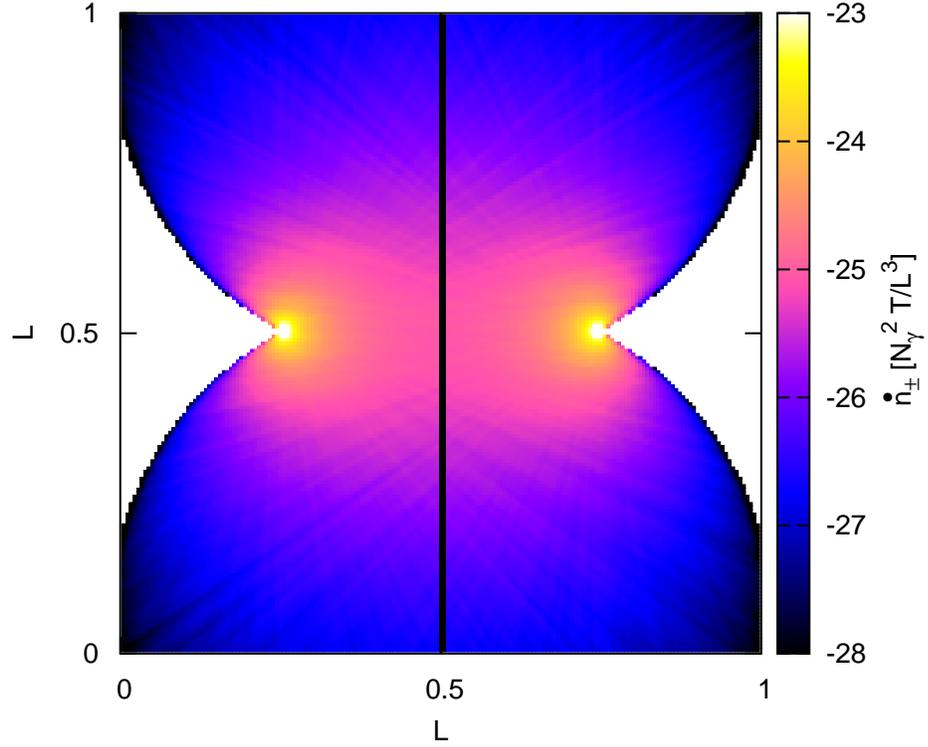}}
\end{picture}
\caption{\normalsize
Test problem: the pair production rate in the plane of two, isotropic point
sources of high energy radiation ($k^t_1=k^t_2=4 m_e c^2$).  Pair production
rate is given here in units of $\dot{N}_{\gamma}^2 T/L^3$, where $L$ is a
length unit, $\dot{N}_{\gamma}$ is a number of photons produced by each source
per unit time T. Pair production rate is zero in two side regions because the
energy of photons in the center-of-momentum frame is below the threshold
energy there. The pair production rate is symmetric with respect to the axis
connecting two sources.  Black contour - see Fig~\ref{fig:test_slice}.
}
\label{fig:test_map}
\end{figure*}

\begin{figure*}
\begin{picture}(0,550)
\put(-120,50){\includegraphics{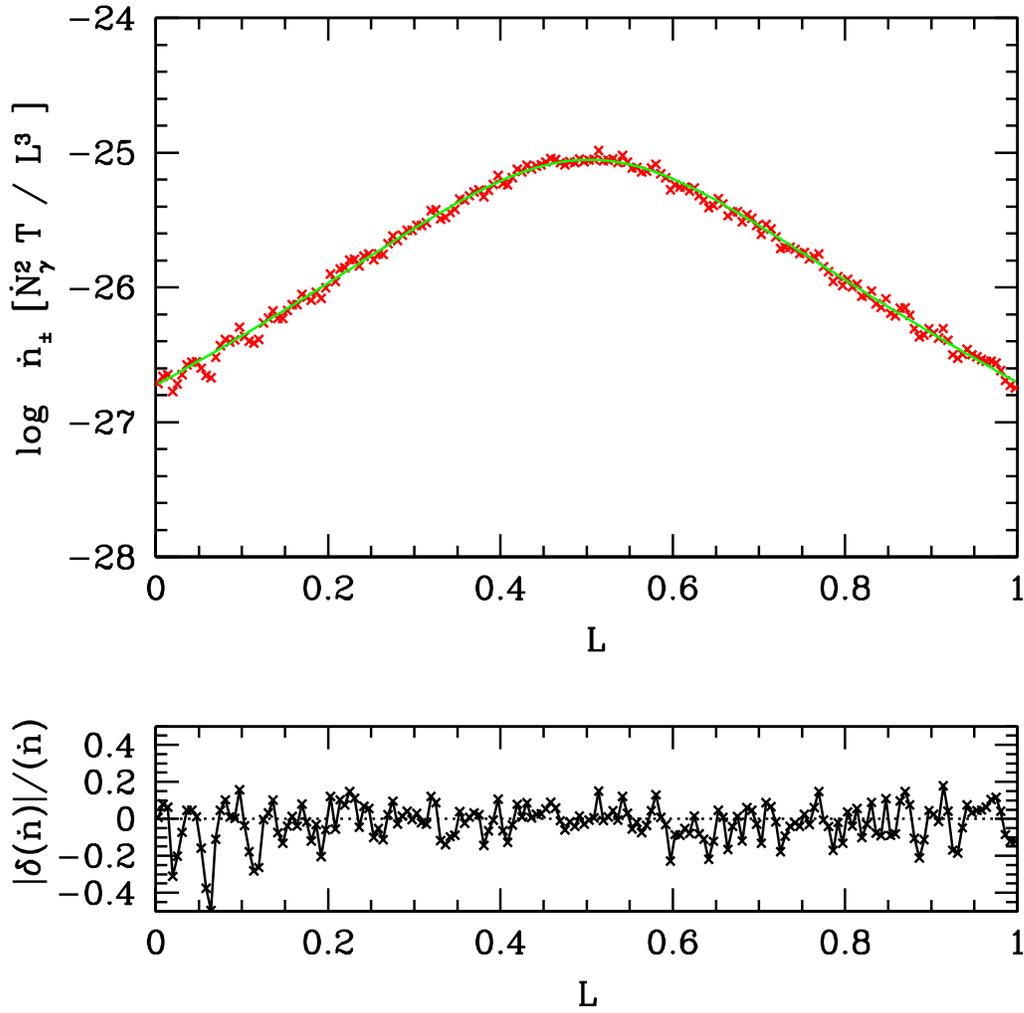}}
\end{picture}
\caption{\normalsize
Test problem: upper panel: Analytical (green line) and numerical (red points)
pair production rates, along the black contour in
Figure~\ref{fig:test_map}. Lower panel: The fractional difference between
analytical and numerical solutions.
}
\label{fig:test_slice}
\end{figure*}

\begin{figure*}
\begin{picture}(0,550)
\put(-120,50){\includegraphics{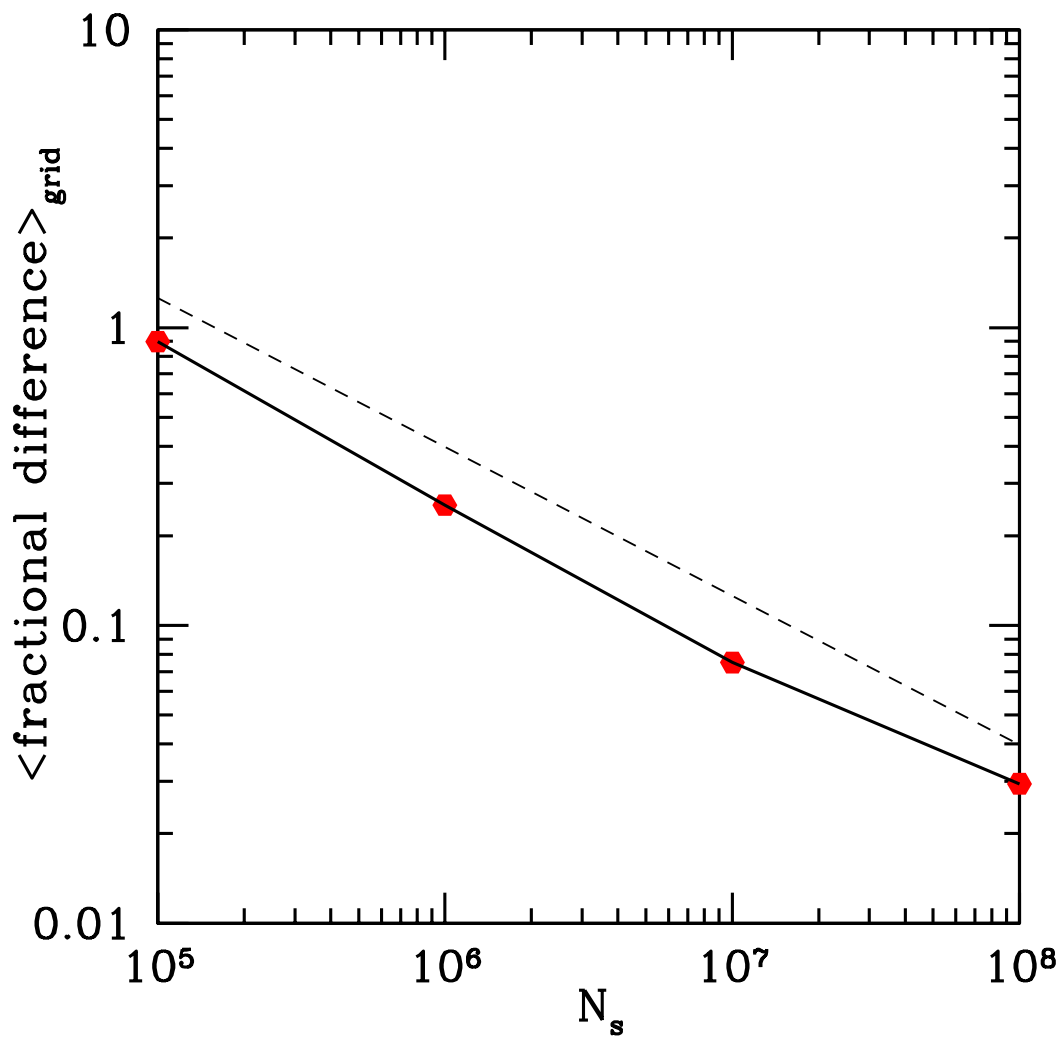}}
\end{picture}
\caption{\normalsize
Test problem: the grid averaged fractional difference between the numerical
and analytical pair production rates as a function of number of photons
packets produced by each source $N_s$. The dashed line is proportional to
$N_s^{-1/2}$. For $N_s=10^7$ the average difference per zone is about 7\%, for
$N_s=10^8$ it is on average less than 3\%.
}
\label{fig:test_convergence}
\end{figure*}

Figure~\ref{fig:test_map} shows a 2D map of the numerically evaluated
pair production rate in the plane of the two sources.
Figure~\ref{fig:test_slice} shows the analytic and numerical pair
production rates along the black contour shown in
Figure~\ref{fig:test_map} (upper panel), and their difference (lower
panel). The error in the numerical rate is $\propto N_s^{-1/2}$, as
demonstrated in Figure~\ref{fig:test_convergence}, where $N_s$ is the
number of photon packets emitted by each source.  Evidently the Monte
Carlo method produces an unbiased, convergent estimate of the pair
production rate.

\section{RIAF scaling laws}\label{pair_scaling}

In this section we derive scaling relations for the pair production rate
in two cases: (1) $M$ and $\dot{M}$ are known and the flow is
radiatively inefficient; (2) the spectrum $\nu L_\nu$ and $M$ are known.
In case (1) we can numerically evaluate the pair production rate
self-consistently and check it against the scaling relation.  In case
(2) we can do the same, but we also obtain a method for estimating pair
production rates from observations that may also apply to flows that are
radiatively efficient.

First consider the pair production rate density at a single point in the flow
where the plasma-frame photon spectrum is a power-law with high energy
cutoff at $\eps = E/(m_e c^2) = \eps_{max} \gg 1$:
\begin{equation}
\frac{d n}{d E} = \frac{n_0}{m_e c^2} \eps^\alpha e^{-\eps/\eps_{max}}
\end{equation}
We evaluated the pair production rate density numerically for this energy
distribution.  A fit to the result over $-3 < \alpha < 2$ and
$10 < \eps_{max} < 160$ gives
\begin{equation}
\frac{\mpprod}{n_0^2 \sigma_T c} \simeq 
\frac{1}{16} e^{2\alpha/3} (\frac{4}{3} + \eps_{max}^{\alpha/2})^4
\ln(\frac{\eps_{max}}{2})
\end{equation}
(Zdziarski [1985] gives a similar expression in the $\eps_{max} \gg
1$ limit).  At worst the fit is $\approx 2$ too small for $\alpha
\approx 0$ and $\eps_{max} = 160$.  For $\alpha < -2$, which is typical
of our models, the relative error is smaller than $60\%$. 

For $\alpha > 0$ ($d\ln \nu L_\nu/d\ln \nu > 2$) pair production is
dominated by photons with $\eps \sim \eps_{max}$, and $\mpprod$ is
therefore sensitive to $\eps_{max}$.  For $\alpha < 0$, pair production
is dominated by pairs with $\eps \sim 1$ in the center-of-momentum
frame.  In this case there is an equal contribution from each
logarithmic interval in energy in the plasma frame, and the pair
production rate density is therefore weakly (logarithmically) dependent
on $\eps_{max}$.

Our models have $\alpha \lesssim -2$ so $\mpprod$ is insensitive to
$\eps_{max}$.  Therefore the effective number density of pair producing
photons is $n_0 \sim L_{512}/(4 \pi \lunit^2 m_e c^3)$, where $L_{512}
\equiv \nu L_\nu(512 {\rm keV})$ and $\mpprod \sim n_0^2 \sigma_T c$.
Then
\begin{equation}
\dot{n}_{\pm} \simeq \left( \frac{ L_{512} }{m_e c^2} \frac{1}{
  {\mathcal L}^2 c} \right)^2 \sigma_{T} c \times f(\frac{r}{\mathcal L},
\mu). \label{general_forumlandot}
\end{equation}
where $f$ is a dimensionless function of Kerr-Schild radius $r$ and 
colatitude $\theta = \cos^{-1}\mu$.

What do we expect for the spatial distribution of pair production $f$?
The pair-producing photons are made by upscattering synchrotron photons
in a ring of hot gas near the innermost stable circular orbit (ISCO).
Away from this ring the density of photons will fall off as $\sim
1/r^2$.  The pair production rate also depends on the angle $\psi$
between photon trajectories in the coordinate frame through the
geometrical factor $\ecm^2/k_1^t k_2^t \propto 1-\cos\psi$.  At large
$r$ $\psi \lesssim 1$ so $1 - \cos\psi \propto 1/r^2$.  Then $\mpprod
\sim r^{-6}$.  Compton upscattered photons are also beamed into the
plane of the disk by the relativistic orbital motion.  If the
intensity of photons to be scattered is nearly independent of $\theta$
then the pair production rate density should fall off away from the
midplane as the density of upscattering electrons $\sim  \exp(-\mu^2/(2
\sigma_{\rho}^2)$ where $\sigma_\rho \simeq 0.3$.  Gathering these
estimates together we expect $f \sim \exp(-\mu^2/(2 \sigma_pm^2)) /
r^6$ where $\sigma_\pm \approx \sigma_\rho$.

\subsection{Scalings with model parameters}

Now suppose we know the mass $M = m_8 M_8$ ($M_8 \equiv 10^8 M_{\odot}$)
and the accretion rate $\dot{M} = \dot{m} \dot{M}_{Edd}$, where
$\dot{M}_{Edd} \equiv L_{Edd}/(\eps_{ref} c^2)$ and $\eps_{ref} = 0.1$
is a reference accretion efficiency.  We assume that photons are
produced in a low frequency synchrotron peak and then scattered to $\sim
512 \keV$ by $n_{sc}$ Compton scatterings, where $n_{sc}$ is $1$ or $2$.

For a plasma that is optically thin to synchrotron absorption at peak,
the total number of synchrotron photons at the peak frequency produced
per unit time is $\dot{N}_{\nu_{peak}} \simeq 4 \pi \nu_{peak}
j_{\nu_{peak}} {\mathcal L^3} / (h\nu_{peak})$, where $j_{\nu_{peak}}$
is the synchrotron emissivity~\footnote{$j_{\nu_{peak}} \simeq 8\sqrt{2}
e^3 n_e B /(27 m_e c^2)$, see \citealt{leung:2010}}. The number density
of synchrotron photons is then $n_{\nu_{peak}} \simeq
\dot{N}_{\nu_{peak}}/(4 \pi {\mathcal L}^2 c)$.  

A fraction $\tau^{n_{sc}}$ of the peak photons are upscattered to $512
\keV$, where $\tau = \sigma_{T} n_e {\mathcal L}$ is the Thomson depth
of the plasma, so $n_{512} = n_{\nu_{peak}} \tau^{n_{sc}}$.  The mean
number of Compton scatterings is $n_{sc}=\log(m_e c^2 / h
\nu_{peak})/\log A$, where $A\approx 16\Theta_e^2$ is the photon energy
enhancement in single scattering by a relativistic electron, so
\begin{equation}
n_{sc} \simeq  a_1 + a_2 \log \frac{m_8}{\dot{m}}.  \label{nsc_risco}
\end{equation}
We determine $a_1$ and $a_2$ numerically but for a reference model with
$\dot{m}=10^{-8}$ and $m_8=4.5 \times 10^{-2}$ (Sgr A*), the average value of
$n_{sc} \approx 1-2$.

Assuming $\dot{M} \sim 4 \pi \rho c {\mathcal L}$, the magnetic pressure
is comparable to the gas pressure and both are $\sim \rho c^2$, the
plasma density, magnetic field strength, and plasma temperature (close
to the virial temperature) scale as
\begin{equation}
n_e \simeq 
\frac{1}{\eps_{ref}} 
\left( \frac{c^2}{G M_8 \sigma_T} \right)
\left(\frac{\dm}{m_8}\right)
\end{equation}
\begin{equation}
\frac{B^2}{8\pi} \simeq 
\frac{1}{\eps_{ref}} 
\left(\frac{m_p c^4} { G M_8 \sigma_T}\right)
\left(\frac{\dm}{m_8}\right)
\end{equation}
and 
\begin{equation}
\Theta_e \simeq \frac{1}{30} \frac{m_p}{m_e}.
\end{equation} 
The mean emission $\Theta_e$ corresponds to the mean value near the
ISCO, and therefore increases with $a_*$.  Combining, 
\begin{equation}
\mpprod \simeq
{\mathcal A} \;
\left( \frac{1}{r_0^3 \tunit} \right) \;
\eps_{ref}^{-(2 n_{sc} + 3)} \alpha_f^2 \; \frac{m_p}{m_e} \;
\dm^{3 + 2 n_{sc}} \;
f(\frac{r}{\mathcal L}, \mu), 
\label{eq:pprod_scaling}
\end{equation}
where ${\mathcal A}$ is a dimensionless constant to be determined 
numerically, $r_0 \equiv$ classical electron radius, and $\alpha_f
\equiv$ the fine structure constant.  From now on unless stated
otherwise we will set $\eps_{ref} = 0.1$ and the mean number of
scatterings $n_{sc} = 1.5$ (numerical results, below, show $1.4 < n_{sc}
< 1.6$ for relevant $M, \dot{M}$).  Then
\begin{equation}
\mpprod \simeq
9 \times 10^{39} \;
{\mathcal A} \;
m_8^{-1}\; \dm^{6} \;
f(\frac{r}{\mathcal L}, \mu).
\label{eq:pprod_scaling2}
\end{equation}

To estimate the jet kinetic luminosity we need the pair production rate:
\begin{equation}
\dot{N}_{\pm} = \int_{r > r_{hor}} \gdet \, d^3x \, \dot{n}_\pm
\end{equation}
where $r_{hor}$ is the horizon radius. Then
\begin{equation}
\dot{N}_{\pm} \simeq 
\mpprod {\mathcal L}^3 \simeq
10^{78} \, {\mathcal A} \,\dot{m}^{3+2n_{sc}} m_8^2 \;\;\; {\rm s^{-1}}
\label{totpairs}
\end{equation}
Only pairs made inside the funnel and at $r > r_{st} \equiv$ stagnation
radius can escape to large radius (they are ``free pairs''); those made
at smaller radius or inside the accretion flow are advected into the
hole.  The free pair fraction is therefore a small multiple of
(\ref{totpairs}).

Evidently the pair production rate density is sensitive to the mass
accretion rate, $\mpprod \sim \dm^6$.  This steep dependence shuts off
pair production at low accretion rates, making it difficult for low
$\dot{m}$ systems like Sgr A* to populate their funnel with pairs.
Below we will show that the implied funnel pair density for Sgr A* falls
below the Goldreich-Julian charge density (see \S\ref{julian}).

\subsection{Scalings with observables}

Assume that from observations we know $M$, the X-ray luminosity $L_X
\equiv l_X L_\odot$ (assuming isotropic emission), and the 
spectral index $\alpha=d\log (\nu L_{\nu})/d \log
\nu$.\footnote{$\alpha$ is used to extrapolate the spectrum from keV
to MeV energies.  It can be evaluated from X-ray data but this can be
inaccurate due to localized spectral features.  It can be evaluated more
accurately from millimeter/X-ray colors.} Self-consistent models then
permit us to calibrate the relation between these quantities and the
pair production rate density.  Since this relation depends only on the
distribution of pair-producing photons within the source, it seems
likely that it can be applied to sources with $\dot{M} >
\dot{M}_{crit}$, in which cooling is important.

If the spectrum is power-law from the keV to MeV energy,
\begin{equation}
L_{512}(L_X) \approx L_X e^{4.92 \alpha}, 
\end{equation}
\begin{equation}
\mpprod \approx 
{\mathcal B} \;
\left(\frac{c^3 \sigma_{T} L_{\odot}^2}{m_e^2 G^4 M_8^4}\right) \;
l_X^{2} \;
e^{(9.26\alpha)} \;
m_8^{-4} \; 
f(\frac{r }{\mathcal L}, \mu) \label{eq:n_obs}
\end{equation}
where ${\mathcal B}$ is a constant to be determined numerically and
$(c^3 \sigma_{T} L_{\odot}^2/m_e^2 G^4 M_8^4) \approx 10^{-8}$
$\rm{cm}^{-3} \rm{s}^{-1}$.  This assumes that the observed spectrum and
the plasma-frame spectrum near the black hole are identical.  We checked
the plasma-frame spectrum and found it to be a slightly blueshifted
version of the observed spectrum; the blueshifting does not change the
scaling relation.

The pair production rate is
\begin{equation} 
\dot{N}_{\pm} \simeq 
\mpprod {\mathcal L}^3 \simeq 
{\mathcal B} \;
(\frac{\sigma_{T} L_{\odot}^2}{ m_e^2 G M_8 c^3}) \;
l_X^2 \;
e^{9.26\alpha} \; 
m_8^{-1} \label{eq:tot_n_obs}
\end{equation}
where $(\sigma_{T} L_{\odot}^2/ m_e^2 G M_8 c^3) = 10^{31}$
$\rm{s}^{-1}$. The dependence on black hole mass changes between
Equations (\ref{eq:n_obs}) and (\ref{eq:tot_n_obs}) because $\lunit
\propto m_8$.

\section{Pair production in RIAF - numerical results}\label{results}

We now evaluate the pair production rate numerically, check whether it
matches the expected scaling laws, and evaluate $f(r,\mu)$.  To do this,
we have run simulations with a range of $M$ and $\dot{M}$, assuming that
the models have equal ion and electron temperatures, $T_e = T_i$.  A
list of model parameters is given in Table~\ref{tab1}.

\subsection{Pair creation rate}\label{mass_deposition}

\subsubsection{Dependence on model parameters: $\dot{m}$, $m$ }
\label{spatialscale}

The $\mpprod$ in models A through H (see Table~\ref{tab1}) is well fit by
\begin{equation} 
\dot{n}_{\pm} (r,\mu) = 3 \times 10^{40} \dot{m}^{3+2n_{sc}} m_8^{-1} \times
(\frac{r}{\mathcal L})^{-6} e^{-\mu^2/(2 \sigma_{\pm}^2)}
\, {\rm cm^{-3}\,s^{-1}}, \label{eq:fit}
\end{equation} 
or ${\mathcal A} \simeq 3$ in equation (\ref{eq:pprod_scaling2}).  The
constant in Equation~\ref{eq:fit} is derived from models with $\Trat=1$.
The constant is sensitive to $\Trat$; for $\Trat = 3$ it is $10^{-4}$
times smaller.  As expected, $\mpprod \sim r^{-6}$ at large $r$;
surprisingly, however, this is also good fit at all $r$.

\begin{figure*}
\begin{picture}(0,550)
\put(-250,0){\includegraphics{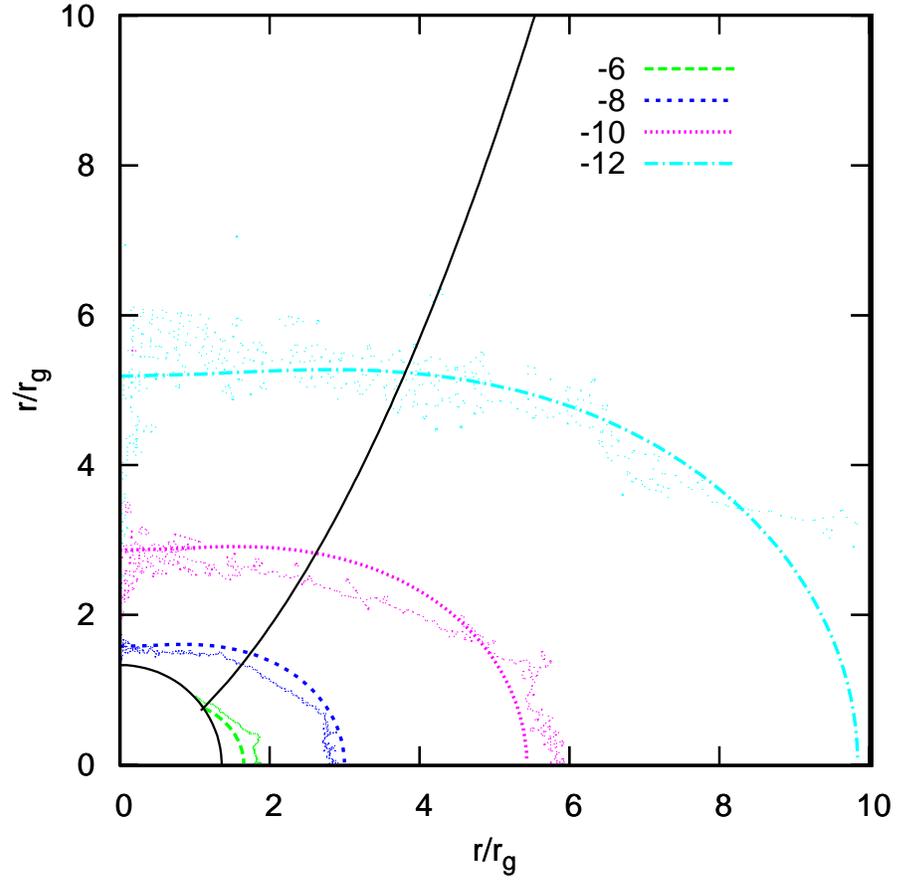}}
\end{picture}
\caption{\normalsize
Spatial distribution of \pprod\ in RIAF model C (points) and the contours of 
corresponding fitting function given by Equation~\ref{eq:fit} (lines). 
The fractional difference between model and data in this case is $< 40\%$.
Black contours mark the black hole horizon and the funnel wall.}
\label{spatial}
\end{figure*}

The pair production scale height $\sigma_{\pm} \approx 0.3$
independent of $\dot{m}$, $m$.  This is nearly identical to
$\sigma_\rho$, the plasma scale height.  Notice that $\sigma_{\pm}$ also
controls $f_{jet}$ the fraction produced inside the funnel.  The funnel wall is at
\begin{equation}
\mu^2 > \mu_f^2 \approx \frac{r + 0.4 G M/c^2}{r + 4 G M/c^2}.
\end{equation}
and $f_{jet} \approx 10\%$.  Figure~\ref{spatial}
shows a 2D contour map of \pprod\ corresponding to model C and
Equation~(\ref{eq:fit}), and a contour marking the approximate funnel
boundary. The grid averaged fractional difference between time averaged
MHD models A through H and Equation~(\ref{eq:fit}) is $ < 60\%$.  Since
$\mpprod$ is a steeply declining function of $\mu^2$, almost all free
pairs are made near the funnel walls.

\begin{figure*}
\begin{picture}(0,550)
\put(-120,50){\includegraphics{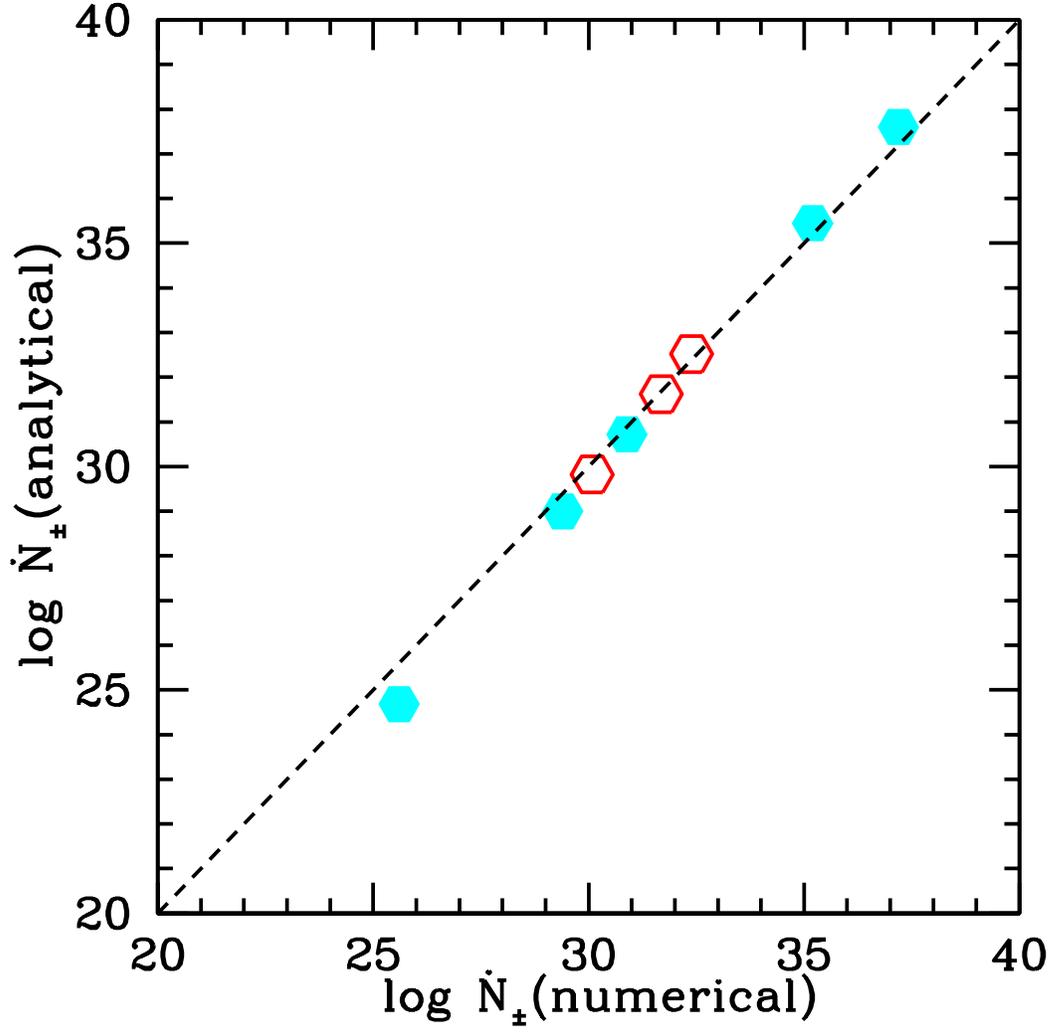}}

\end{picture}
\caption{\normalsize
Pair production rate dependence on the model parameters.  Comparison of the
total pair production rate $\dot{N}_{\pm}$ to the fitting formula for models
with various mass accretion rates $\dot{m}$ (A-E, blue filled symbols), and 
black hole masses $m$ (F-H, red open symbols).
The $\dot{N}_{\pm}({\rm analytical})$ is given by
Equation~\ref{eq:fit_int}.  
}
\label{fig:ABC_dn_mdot}
\end{figure*}

The pair production rate is well fit by
\begin{equation}
\dot{N}_{\pm} = 4\times10^{80}\, \dot{m}^{3+2n_{sc}} \, m_8^{2} \,\,  
{\rm s^{-1}} 
\label{eq:fit_int}
\end{equation}
where a fit gives 
\begin{equation}
n_{sc} = 1 + 0.03 \ln(m_8/\dm).
\end{equation}
Figure~\ref{fig:ABC_dn_mdot} compares the time-averaged numerical
$\dot{N}_{\pm}$ to Equation~\ref{eq:fit_int}.

\subsubsection{Dependence on $l_X$, $\alpha$, $m$}
\label{scale_with_Lx}

The self-consistent radiative model enables us to calculate the emergent
spectrum, from which we can measure a $2-10$keV luminosity $l_X$ and a
spectral slope $\alpha$.  The pair production rate density can be
measured in the same models.  The numerical results are well fit by
\begin{equation} 
\dot{n}_{\pm} (r,\mu) =  10^{-8} \, l_X^{2} e^{9.26 \alpha} m_8^{-4} \times
(\frac{r}{\mathcal L})^{-6} e^{-\mu^2/2 \sigma_{\pm}^2}
\;\; {\rm cm^{-3}\,s^{-1}}. 
\label{eq:fit2}
\end{equation} 
or ${\mathcal B} \simeq 1$ in Equation (\ref{eq:n_obs}).  The fractional
error of the fit is $ < 50 \%$ for time averaged models A-L.  

\begin{figure*}
\begin{picture}(0,550)
\put(-120,50){\includegraphics{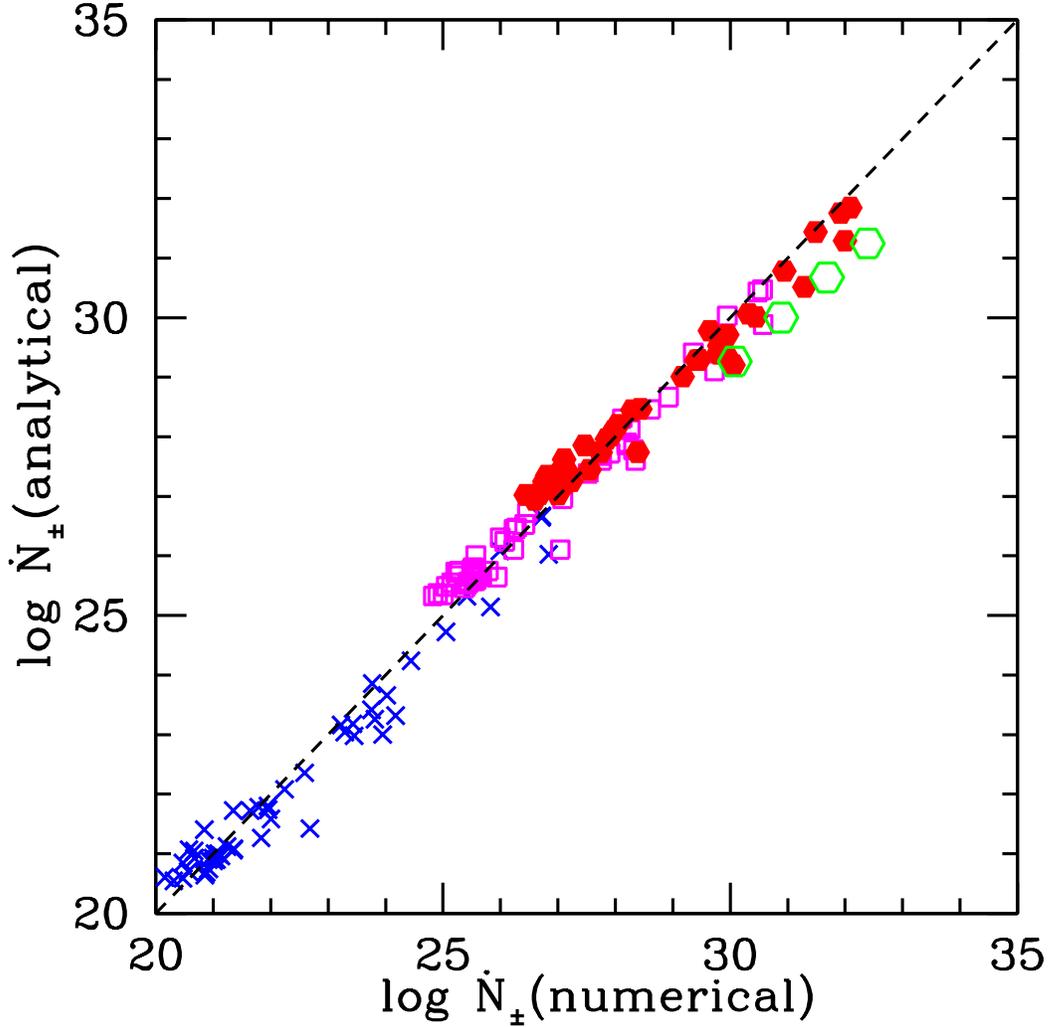}}
\end{picture}
\caption{\normalsize
Pair production rate dependence on observable parameters. Comparison of the
total pair production rate $\dot{N}_{\pm}$ to the fitting formula for models
with different X-ray luminosities $(\nu L_{\nu})_{2-10 keV}$, X-ray spectral
index $\alpha$ and masses.
Crosses, open squares, filled circles 
correspond to different snapshots in models A, B, and C, respectively. Open
circles mark time averaged data from models with different masses (F, G, and
H). The $\dot{N}_{\pm} ({\rm analytical})$ is given by
Equation~\ref{eq:fit_int_lx}.
}
\label{fig:ABC_dn_LX}
\end{figure*}

The pair creation rate is well fit by
\begin{equation}
\dot{N}_{\pm} = 5 \times 10^{30} \, l_X^2 e^{9.26\alpha} m_8^{-1} \;\; 
{\rm s^{-1}} \label{eq:fit_int_lx}
\end{equation}
Figure~\ref{fig:ABC_dn_LX} compares $\dot{N}_{\pm}$ to the semianalytic
formula given by Equation~\ref{eq:fit_int_lx} for different snapshots of
the simulations with different mass accretion rates (models A-C) and
black hole masses (models F-H). The semianalytic and numerical results
agree well, and the scaling constants are close to those estimated in \S
\ref{pair_scaling}.

Although Equations (\ref{eq:fit2}) and (\ref{eq:fit_int_lx}) are
strictly valid only for radiatively inefficient flows with $\dot{m} <
\dot{m}_{crit}$, they depend mainly on the geometry of the radiation
field and not on the radiative efficiency of the flow.  We speculate
that they provide a good estimate of the pair production rate even in
more efficient systems, if $\sigma_\pm$ is set to the scale height of
the Comptonizing corona and the spectrum extends to sufficiently high
energy.

\subsection{Pair power and electromagnetic luminosity of a funnel}

We define the funnel pair creation ``luminosity'' 
\begin{equation}
L_{\pm} \equiv f_{jet} \, \dot{N}_\pm \, 2 m_e c^2 \Gamma_{jet}
\end{equation}
where $\Gamma_{jet}$ is the jet bulk Lorentz factor at large $r$
(assuming cold flow).  Then
\begin{equation}
L_{\pm} \simeq   6 \times 10^{74} f_{jet} \, \dot{m}^{3+2 n_{sc}} m_8^{2}
\Gamma_{jet}
\, \ergps 
\end{equation}
and 
\begin{equation}
L_{\pm} \simeq 10^{25}  f_{jet} \, l_X^2 e^{9.26\alpha} m_8^{-1} 
\Gamma_{jet} \, \ergps.
\end{equation}
It is interesting to compare this with the Blandford-Znajek (BZ), or
electromagnetic, luminosity of the funnel
\begin{equation}
L_{BZ}= 2 \pi \int_{\mu^2 > \mu_f^2} d\theta \sqrt{-g} T^r_t
\label{eq:bz}
\end{equation}
where $T^r_t = b^2 u^r u_t - b^r b_t$ is the electromagnetic part of the
stress-energy tensor and is computed directly from the simulation data.
The BZ luminosity is well fit by
\begin{equation}
L_{BZ} \approx 8 \times 10^{45} 
(1-\sqrt{1-a_*^2})^2 \, \dot{m} \, m_8 \;\; \ergps
\end{equation} 
The scaling with $a_*$ is taken from Equation (61) of
\citealt{mckinney:2004}, which is a fit to numerical data.

For mass comparable to that of Sgr A* and $a_* = 0.94$, $L_{BZ} >
L_\pm/\Gamma_{jet}$ for $\dot{m} < \dot{m}_{crit} \approx 10^{-6}$ (for
$\dot{m}_{crit}$ see \S~\ref{selfcon}).  At low accretion rates the BZ
luminosity completely dominates the pair luminosity, because the pair
luminosity is such a steep function of accretion rate.  Because $L_\pm
\sim m_8^2$ while $L_{BZ} \sim m_8$, the $\dot{m}$ at which $L_{BZ} \sim
L_\pm$ is higher for lower mass black holes.  $L_{\pm}/L_{BZ}$ ratio is
shown in the Table~\ref{tab1}.  For reasonable $\Gamma_{jet}$ the funnel
luminosity is therefore electromagnetically dominated for radiatively
inefficient flows. 

\subsection{Energy-momentum deposition}

Some of the pairs created in the funnel will escape to large radius and
some will fall into the black hole. In an MHD model, the escaping
fraction and asymptotic Lorentz factor will depend on the run of pair
creation rate with radius, the magnetic field structure, the energy
density of the pair plasma, and the pair-creation four-force $G^\mu$.

Based on numerical calculations the spatial distribution of $G^\mu$ is
well fit, with $x \equiv r/\lunit$, by
\begin{equation}
G^{0}_{code}(r,\mu) = G^0  \frac{{\mathcal L^2}{\mathcal T^2}}{\mathcal M} 
\approx  
\frac{300}{x} \; 
\frac{\mpprod(x,\mu) m_e c}{\munit/(\lunit^2 \tunit^2)}
\end{equation}
\begin{equation}
G^{1}_{code}(x,\mu) = G^1  \frac{{\mathcal L^2}{\mathcal T^2}}{\mathcal M} 
\approx 
\frac{20 (x-x_{st})}{x^2} \;
\frac{\mpprod(x,\mu) m_e c}{\munit/(\lunit^2 \tunit^2)}
\end{equation}
\begin{equation}
G^{2}_{code}(x,\mu) = G^2  \frac{{\mathcal L^2}{\mathcal T^2}}{\mathcal M} 
\approx 
\frac{\mu}{x^2} \;
\frac{\mpprod(x,\mu) m_e c}{\munit/(\lunit^2 \tunit^2)}
\end{equation}
\begin{equation}
G^{3}_{code}(x,\mu) = G^3  \frac{{\mathcal L^2}{\mathcal T^2}}{\mathcal M} 
\approx 
\frac{150}{x^2} \;
\frac{\mpprod(x,\mu) m_e c}{\munit/(\lunit^2 \tunit^2)}
\end{equation}
where $\munit, \tunit, \lunit$, and $\mpprod$ are given in cgs units.
The four-force vector components are given in a Kerr-Schild coordinate
basis and in code units; we divide $G^{\mu}$ in cgs units by the
unit of the four-force density ${\mathcal M}/{\mathcal L^2}{\mathcal T^2}$.
All components of the four-force depend more steeply on
radius than \pprod.  The radial component of the four-force is positive
at large radius, zero at $x_{st} \approx x_{ISCO} (1 + \mu^2/2)$, and
negative at small radius.  The sign of $G^{2}$ changes at the equatorial
plane, as it should.

Pairs are created in the funnel with an initial distribution function,
which is immediately isotropized with respect to rotation around the
magnetic field.  Later evolution of the distribution function depends on
ill-understood relaxation processes;  the pairs may not relax. Whether
or not relaxation occurs the initial mean energy of the particles is of
interest.  So: what is $d\dot{n}_{\pm}/d\log \gamma_{e^\pm [FF]}$
($\equiv$ the energy distribution of newly created pairs), where
$\gamma_{e^\pm [FF]}$ is the Lorentz factor of new particles measured in
the fluid frame [FF]?  

\begin{figure*}
\begin{picture}(0,550)
\put(-120,50){\includegraphics{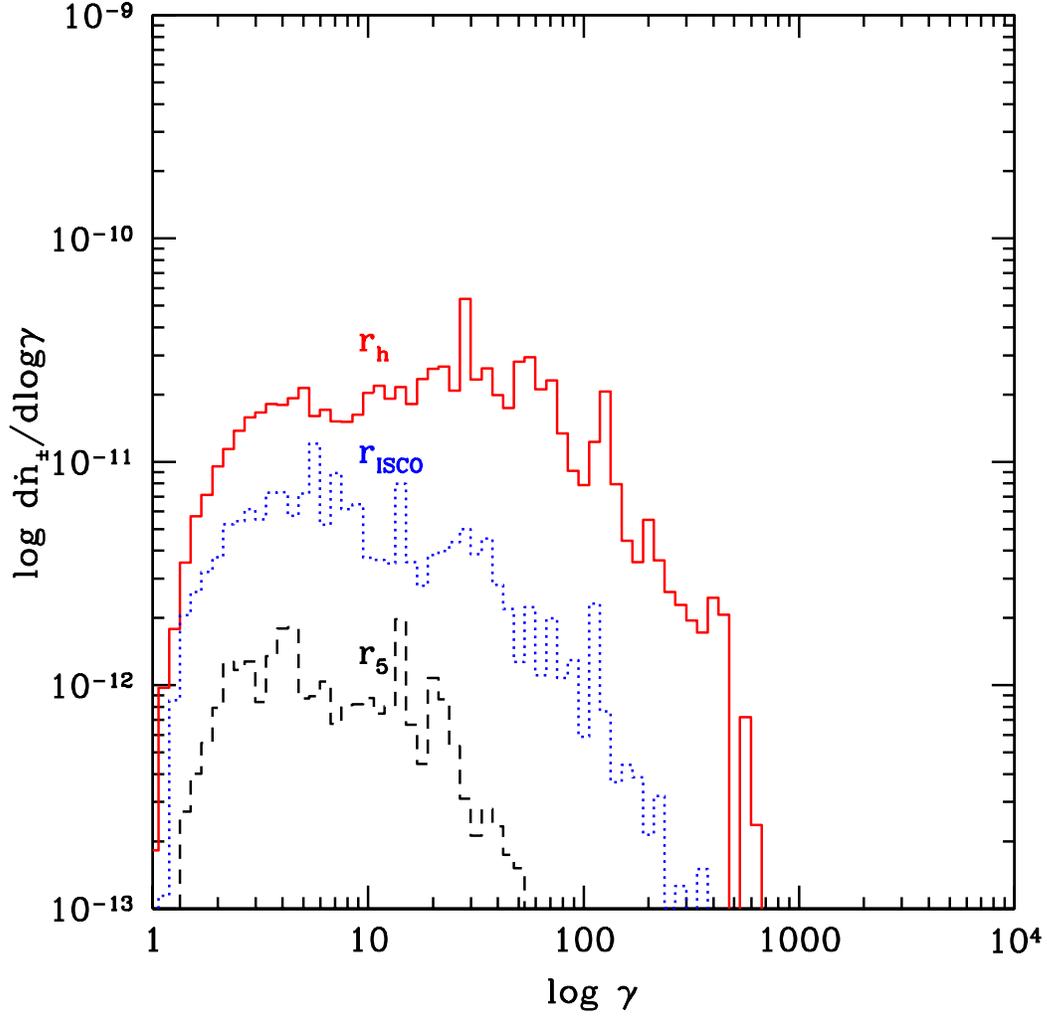}}
\end{picture}
\caption{ \normalsize
The energy distribution of $\mpair$ pairs produced in the magnetized funnel where
$\gamma=\gamma_{e^\pm [FF]}$ is measured in the plasma frame at different
radii (single time slice of model C).}
\label{fig:gamma_slope_jet}
\end{figure*}

Four-momentum is conserved in pair creation, so the average Lorentz
factor of the new leptons in the fluid frame is
\begin{equation}
\gamma_{e^\pm [FF]} = -\frac{1}{2} u_{\mu} (k^\mu+k'^\mu)
\end{equation}
where $u_{\mu}$ is four-velocity of the background plasma and we assume
that $k^\mu$ is in units of $m_e c^2$.
Figure~\ref{fig:gamma_slope_jet} shows $d\dot{n}_\pm/d\log\gamma_{e^\pm
[FF]}$ in model C at $r=r_h, r_{isco}, 5 \Rg$
(averaged over 20 radial zones from $\theta=0-10\deg$; here $r_h \equiv$ 
event horizon radius).  The distribution is flat and cuts off at
$\gamma_{e^\pm [FF] max} \sim 100$ at $r=r_{isco}$. 
 
If the thermalization timescale is short (which, given the low density
and likely high temperature of the plasma, seems unlikely to us), then
the rate of internal energy injection due to $\mpair$ creation is
$\dot{u}=\dot{e}_{\pm}$, where the kinetic energy density injection
rate, in the plasma frame, is \footnote{$\dot{e}_{\pm}$ can be
obtained by transforming $G^\mu$ from coordinate to fluid frame
and subtracting the rest mass energy.}
\begin{equation}
\dot{e}^{\pm}= \int d\gamma_{e^\pm [FF]} \,
(\gamma_{e^\pm [FF]}-1) \, m_e c^2 \,
\frac{d \dot{n}_{\pm}}{d \gamma_{e^\pm [FF]}} 
\end{equation}
The corresponding temperature of the newly injected pairs is
$\Theta_{e^\pm} = (1/3) \dot{e}_{\pm}/(\dot{n}_{\pm} m_e c^2) \approx
(20,10,3)$ at $r=(r_h, r_{isco}, 5 \Rg)$.  This temperature is
comparable to the thermal background plasma and is sub-virial. Pairs are
not born hot.

The entropy of the injected pairs is likely to increase over the initial
entropy.  The funnel is exposed to ``acoustic'' radiation from the
accretion disk.  A small fraction of the MHD waves generated by
turbulence in the disk will propagate toward the funnel, be transmitted
at the funnel wall, and dissipate within the funnel.  Because the
density of the funnel is so low, even a small fraction of the disk
acoustic luminosity is capable of raising the pair plasma temperature to
$\Theta_e \gg 1$ before it can escape at Lorentz factor $\Gamma_{jet}
\gg 1$.  Until the magnitude of turbulent heating can be estimated, the
dynamics of the funnel pair plasma is extremely uncertain.

\subsection{Comparison to Goldreich-Julian density}\label{julian}

Pairs are created and then are accreted or escape on the light-crossing
time $~ \tunit$.  This implies a density $n_\pm \sim \mpprod \tunit$ in
the funnel.  Is $n_\pm$ sufficient to enforce the ideal MHD condition
${\bf E} = 0$ in the rest frame of the plasma ($u_\mu F^{\mu\nu} = 0$):
does the pair number density exceed the Goldreich-Julian density
$n_{GJ}$?  

A naive estimate of $n_{GJ}$ uses the flat-space Goldreich-Julian charge
number density
\begin{equation}
n_{GJ} \simeq \frac{\Omega B}{4\pi e c} = \frac{a_* B c^2}{32 \pi G M e}
\end{equation}
where the field rotation frequency in the funnel $\Omega = (a_*/8)
c^3/(G M)$ in the Blandford-Znajek model at $a_* \lesssim 1$.  For our
standard Sgr A* model with $B \sim 30$G, $a_* \simeq 0.94$, $M \simeq 4.5
\times 10^6 \MSUN$, so $n_{GJ} \simeq 10^{-3} \cm^{-3}$.

A better estimate uses the Blandford-Znajek model for a monopole
magnetosphere.  We use Kerr-Schild coordinates $(t,r,\theta,\phi)$ and
define the Goldreich-Julian density as the charge density measured in
the frame of the normal observer, who to lowest (zeroth) order in $a_*$
has four-velocity $n_\mu = (-(1 + 2/r)^{-1/2}, 0, 0, 0)$.  Using the
current density $J^\mu$ derived from the BZ monopole solution as given
in \cite{mckinney:2004} we find, to lowest order in $a_*$,
\begin{equation}
n'_{GJ} \equiv -n_\mu J^\mu = 
\frac{a_* B^r c^2}{4 \pi G M e} \;
\frac{(1 + 2/x)^{1/2} \cos\theta}{x^3}
\end{equation}
where $x \equiv r/\lunit$ and $B^r$ is the radial component of the field
at $x = 1$.  This is close to the naive estimate.  \footnote{Since
$n'_{GJ} \sim B^r \cos\theta$ the charge density changes sign from one
hemisphere to the other.  In a split monopole model the sign would be
the same.}  Notice that $n'_{GJ} \sim 1/r^3$, so $n_\pm/n_{GJ}$ is
smallest closest to the horizon if $n_\pm \sim 1/r^2$ as, for example,
in a wind.

A still better estimate for $n_{GJ}$ would use the simulation-derived
currents.  We have checked these and the charge densities are consistent
with the estimates just given.  

Using estimates for $B^r$ from \S~\ref{pair_scaling}, we can derive a
scaling with $\dot{m}$ and $m$:
\begin{equation}
n_{GJ} \sim
3 \times 10^{-1} \;
a_* \; 
\dot{m}^{1/2} \; m_8^{-3/2} \;\; \cm^{-3}
\end{equation}
near $x=2$.  Then using $n_{\pm} \sim \dot{n}_\pm
\tunit \approx 5\times 10^{40} \dot{m}^6$ (using Equation~\ref{eq:fit} and
$\mu^2=1$):
\begin{equation}
\frac{n_{GJ}}{n_{\pm}} \sim 6 \times 10^{-42} \;
a_* \; \dot{m}^{-11/2} \; m_8^{-3/2}.
\end{equation}
This is the ratio at the axis.  Although the Goldreich-Julian density
varies only weakly across the funnel, $\mpprod(\mu^2 =
\mu_f^2)/\mpprod(\mu^2 = 1) \simeq 20$ at $x \sim 2$.  At very low
$\dot{m}$, where $n_{GJ} \sim n_\pm$, the center of the funnel is
populated by some other process---perhaps a pair cascade--- and the
edges by pair production in $\gamma\gamma$ collisions.

\section{Discussion}\label{discussion}

\subsection{Selfconsistency of the models}\label{selfcon}

\begin{figure*}
\begin{picture}(0,550)
\put(-120,50){\includegraphics{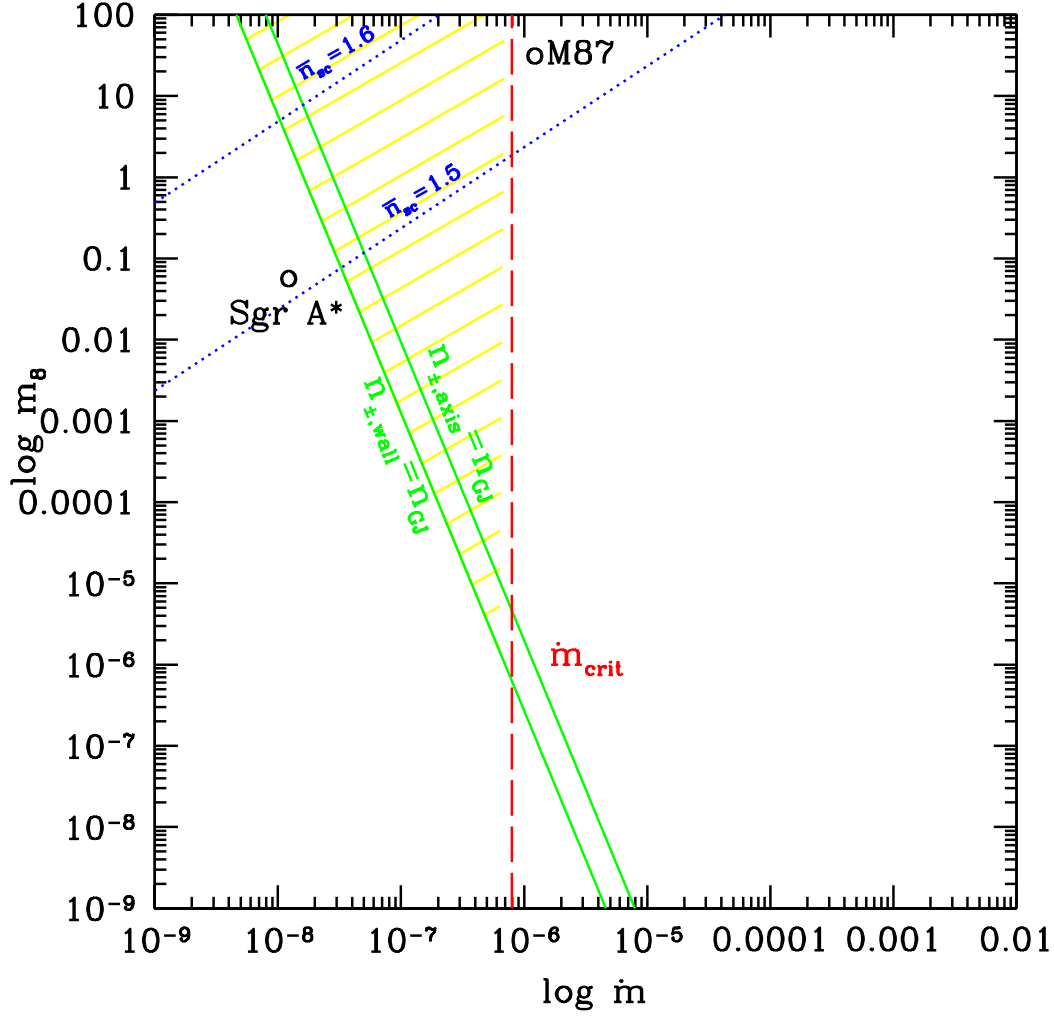}}
\end{picture}
\caption{ \normalsize
Fitting formulas are fully selfconsistent with the model assumptions
for $m$ and $\dot{m}$ within the shaded region. 
Two solid lines mark regions where the $n_{GJ}$ equals to the pair density at
the funnel axis and funnel wall. Dotted lines shows scaling law
for the number of scatterings ($n_{sc}$).
\sgra\ and M87 are marked as open circles. 
}
\label{last_fig}
\end{figure*}

Our models are self-consistent when they are radiatively inefficient and
$n_\pm$ is greater than the Goldreich-Julian density.
Figure~\ref{last_fig} shows the self-consistent $\dot{m}, M$ as a shaded
region.  The region is bounded at low accretion rates by the solid line,
where $n_{\pm}=n_{GJ}$ (for $a_*=0.94$).  The region is bounded at high
accretion rates by $\dot{m} = \dot{m}_{crit}$ (vertical dashed line),
where the model becomes radiatively efficient (here defined as
$L_{Bol}/\dot{M}c^2 = 0.1$).  Numerically, $\dot{m}_{crit} \approx
10^{-6}$ at $\Trat=1$.  The models are fully self-consistent in the
resulting wedge in parameter space.  They are never applicable to
stellar mass black holes, which cannot produce enough pairs to exceed
the Goldreich-Julian density even at $\dot{m}_{crit}$.

\subsection{\sgra}

The bright radio source associated with the $M = 4.5 \times 10^{6}
\MSUN$ black hole in the Galactic center, \sgra, is a weak X-ray source
(`quiescent' emission $L_X \lesssim 10^{33} \ergps$) and is strongly
sub-Eddington ($L/L_{Edd} \approx 10^{-9}$).  \cite{moscibrodzka:2009}
have presented models of \sgra\ that suggest the most probable spin of
the black hole $a_*=0.94$, temperature ratio $T_i/T_e = 3$, and
$\dot{m}=2 \times 10^{-8}$. The $\mpprod$ rate for these parameters is
lower than the scaling laws of \S 5 because $\Trat > 1$.  For the
purposes of this subsection only we consider 3D GRMHD models that have
$\mpprod$ very similar to the 2D models.  All models assume the
accretion flow lies in the equatorial plane of the black hole. 

During the quiescent state the X-ray luminosity is $l_X < 1$.  Near the
horizon in the funnel, in model I, $\dot{n}_{\pm} \approx 10^{-9} \,
{\rm s}^{-1}$.  The light crossing time is $\tunit \approx 20 {\rm s}$,
so a typical pair density near the horizon in the funnel is ${n}_{\pm}
\approx 10^{-8} {\rm cm^{-3}}$.  This is five orders of magnitude below
$n_{GJ}\approx 10^{-3} {\rm cm^{-3}}$.  In quiescence the funnel must
therefore (for the assumed spin) be populated by a process other than
$\gamma\gamma$ pair production process considered here, for example a
pair cascade.

\sgra\ exhibits intraday variability at all observed wavelengths (radio,
sub-mm, NIR, and X-rays).  In particular in 2-10 keV luminosity may
increase up to 160 times (the brightest flare detected has luminosity of
$L_X=5.4 \times 10^{34}$ \citealt{porquet:2008}) from the `quiescent'
level and last a few ks.  During a bright flare the X-ray slope can
change, and the implied pair density may reach or exceed $n_{GJ}$.  Even
during a flare the funnel kinetic luminosity is far below the
Blandford-Znajek luminosity (see Table 1) for any reasonable
$\Gamma_{jet}$.  We conclude that close to the black hole any jet in
\sgra\ is electromagnetically dominated.

\subsection{M87}\label{M87_dis}

The core of M87 hosts a sub-Eddington black hole with $M = 3 \times 10^9
\MSUN$ (\citealt{marconi:1997}, but see \citealt{gebhardt:2009}) at
distance $D \simeq 16$Mpc. M87 has a prominent radio jet resolved from
$100 GM/c^2 - 1$kpc.  \cite{reynolds:1996} have argued that models in
which the jet is made of a pair plasma are favored over those in which
the jet is composed of an ion-electron plasma.  It is difficult to apply
our model to M87 because the SED of M87 from $r < 100 \Rg$ includes
contributions from both the accretion flow and the jet.

We assume $a_*$=0.94, set the inclination $i=30\deg$ \citep{heinz:1997},
assume that the accretion disk lies in the equatorial plane of the black
hole, that the electron distribution function is thermal, and that
$\Trat = 3$. The SED is normalized via $f_\nu(\nu = 230 \GHz) = 1770
{\rm \, mJy}$ at $16$ Mpc \citep{tan:2008}. We find that the resulting
zero cooling, 2D model with $\dot{m}=1.5 \times 10^{-6}$ (model K) is
radiatively efficient (see Table 1) and therefore not self-consistent.

\begin{figure*}
\begin{picture}(0,550)
\put(-120,50){\includegraphics{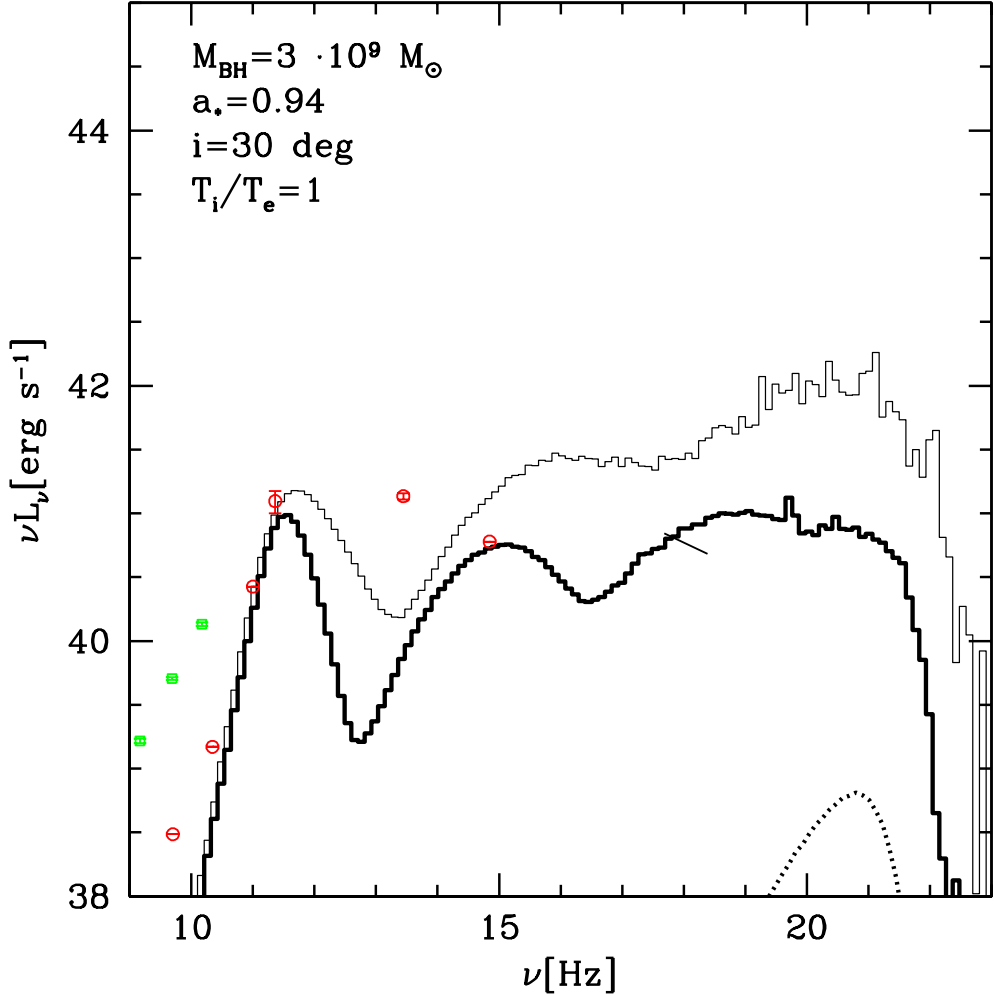}}
\end{picture}
\caption{ \normalsize
Model L: time averaged spectral energy distribution.
Two lines show model with $\dot{m}=10^{-6}$ and $\Trat=1$.
$\dot{m}$ is chosen to normalize to 1.7 Jy at 230 GHz.
Thick-solid line corresponds to run in which a radiative cooling is taken into
account as described in the Appendix. Dotted lines is a free-free process spectrum.
Model, that does not accounts for any cooling in the MHD simulation, is marked
as thin-solid line (also this model is not shown in the table). 
Observational points are taken from:
\citet{reynolds2:1996},
\citet{tan:2008} (230 GHz), 
\citet{perlman:2001} (10.8 $\mu m$),
\citet{harms:1994} ( $7 \times 10^{14}$ Hz), 
\citet{dimatteo:2003} (2-10 keV) }
\label{M87_SED_L}
\end{figure*}

To find a more self-consistent model we have run a GRMHD simulation with
 synchrotron and bremsstrahlung cooling, but not Compton
cooling, included (model L, with $\dot{m} = 10^{-6}$).  The cooling
rates and cooling algorithm are presented in the Appendix. The
efficiency is reduced to $\sim 30\%$.
Figure~\ref{M87_SED_L} shows the SED for model L; the dotted line is the
bremsstrahlung contribution, which is negligible.  The model has $L_X =
10^{41} \ergps$, $L_{\pm}=7 \times 10^{36} \ergps$, and $L_{BZ} \approx
10^{41} \ergps$.  

If we identify $L_{BZ}$ as the jet luminosity then the model is
inconsistent with existing estimates (see the useful compilation of
estimates in Table 3 of \cite{li:2009}), which range from $3 \times
10^{42} \ergps$ \citep{young:2002} to $> 10^{44} \ergps$ estimated by
\cite{bicknell:1996}. 
The discrepancy between $L_{BZ}$ in model L and observations 
is by 1-3 orders of magnitude, but the lowest 'observed' value of $L_{BZ}$ 
could be possibly reached in a model which combines $\Trat \gtrsim 1$ and
radiative cooling.

The jet is optically thin to pair annhilation: $\tau_\pm
\approx n_\pm {\mathcal L} \sigma_T \approx 10^{-10}$.
It is also optically thick to pair production for TeV photons, 
$\tau_{\gamma\gamma} (r) \sim \sigma_T n_{IR}\lunit \sim 10^{3}$, 
($n_{IR} \approx 10^{13} {\rm cm^{-3}}$ 
is the infrared photon density calculated from the Monte Carlo simulations).

The shape of the spatial distribution of pair production in model L is
similar to that in models without cooling (although the scaling of the
distribution changes).  The implied pair density $n_{\pm} = \dot{n}_\pm
\tunit \approx 10 \, {\rm cm^{-3}}$ ($\tunit \approx 10^4$ s), which is
$10^{7}$ times larger than $n_{GJ} \approx 10^{-6} {\rm cm^{-3}}$ in
almost the entire computational domain.  

Because of the shortcomings of the model, however, it is useful to use a
more nearly model-independent estimate of the total pair production rate
based on Equation (\ref{eq:fit_int_lx}).  For $L_X \simeq 3 \times
10^{41}$ ($7 \times 10^{40}$ from \cite{dimatteo:2003} corrected upward
to an isotropic X-ray luminosity because our models beam X-rays into the
equatorial plane) and $\alpha_{X} = 0$,  $\dot{N}_{\pm} \simeq 10^{45}
{\rm s}^{-1}$.  This implies $L_K = f_{jet} \dot{N}_\pm m_e c^2
\Gamma_{jet} = 8 \times 10^{38} \Gamma_{jet} f_{jet}$.  The implied pair
density exceeds $n_{GJ}$ for model L by $\sim 10^8$.  Since $n_{GJ}
\propto B \propto \dot{m}^{1/2}$ and $n_\pm \propto L_{512}^2$ the
implied pair density will fall below the Goldreich-Julian density only
for $(\dot{m}/10^{-6})^{1/2} (L_{512}/10^{41.5})^{-2} < 10^8$.  Even if
$\dot{m} \sim 10^{-4}$ this would require $L_{512} \sim 10^{38}$, which
seems implausibly low given the $\sim 10^{40} \ergps$ TeV luminosity
\citep{aharonian:2006}.
Therefore the main conclusion of this section does not change even if 
a more selfconsistent model is found.

There are significant limitations on the model.  We have considered only
one value of $a_*$; estimates and preliminary models not described here
show that the pair production rate is a steeply increasing function of
$a_*$.  Further preliminary models and a comparison of the $\Trat = 3$
model for \sgra\ with the scaling relation for $\Trat = 1$ models also
show that the pair production rate declines sharply as $\Trat$
increases.  But the allowed values of $\Trat$ are strongly constrained
by submm VLBI \citep{fish:2010}, because as $\Trat$ increases so does
the size of the synchrotron photosphere.  

After submission of this article \citet{levinson:2010} released a paper
focused on modeling TeV emission and pair production in M87 (and Sgr
A*). These authors use an ADAF model, assume that $T_e$ saturates at $~
few \times 10^9 K$ ($\Theta_e \sim 1 $), and set $\dot{m}\approx
10^{-4}$. The model is semi-analytic and does not include general
relativistic effects.  Bremsstrahlung is the dominant source of photons
near the pair-production threshold, and the resulting radiation field is
inadequate to raise the pair density above $n_{GJ}$.
\citet{levinson:2010} therefore invoke a gap/pair cascade model to
produce pairs.  

We have investigated the \cite{levinson:2010} model by calculating
images and an SED for a GRMHD/radiative transfer model with $\Theta_e=1$
everywhere, $\dot{m} = 10^{-4}$, and $a_*=0.94$.  The model includes
synchrotron, Compton and bremsstrahlung.  We find $f_{230GHz}=1 Jy$ (at
$i=30 \deg$), and $L_{BZ} =10^{43} \ergps$, consistent with
observations.  Free-free cooling dominates over synchrotron cooling only
at $r>20 \Rg$.  Levinson \& Rieger neglect Compton cooling, but we find
that Compton $y = A \tau \approx 12$ and that with Compton cooling
included the model efficiency is $\approx 200\%$.   The parameter space
is large and the spectrum is parameter-sensitive, so there may be nearby
models (with different $a_*$, $\Theta_e$, $\dot{m}$, $i$) that are
radiatively inefficient.  The main point, however, is that
self-consistent models can contain surprises that might not be
anticipated in quasi-analytic estimates.  Comptonization, in particular,
occurs close to the innermost stable circular orbit, is therefore
sensitive to the spin, and requires proper treatment of gravitational
lensing.  We concur with Levinson \& Rieger's conclusion that the pair
production rate due to $\gamma\gamma$ collisions is small.

The model is also constrained by VLBI measurements.  An optically thick
spherical source of radius $r$ and distance $D$ in the Rayleigh-Jeans
regime has flux $f_\nu \approx 2 \pi \Theta_e m_e c^2 (r/D)^2/\lambda^2$.
Small $r$ inferred from VLBI therefore requires high $\Theta_e$.  At 230
GHz \citep{fish:2010} report structure on scales of a ``few
Schwarzschild'' radii, while we find the Levinson \& Rieger model has a
photosphere at $\approx 30 GM/c^2$.  In comparison, our model L has a
photosphere at $\sim 7 G M/c^2$.  This argues against the Levinson \&
Rieger model if the reported structure arises from the accretion flow
rather than the jet.

\section{Summary}\label{summary}

We have studied electron-positron pair production in black hole
magnetospheres by $\gamma\gamma$ collisions. Our pair production rate
simulations are based on a GRMHD time dependent model of a magnetized
disk around a spinning black hole. The disk is a source of high energy
radiation formed in multiple Compton scatterings of synchrotron photons.
The pair production rates are calculated nearly {\it ab-initio} within
$40 G M/c^2$ of the event horizon, using Monte Carlo methods.

The main results of this work are the fitting formulae for the rate and
spatial distribution of pair production in terms of $m_8$ and $\dot{m}$
(Equation [\ref{eq:fit}]) and in terms of $m_8$, $L_X$, and $\alpha$
(Equation [\ref{eq:fit2}]).  These indicate that $\gamma\gamma$ pair
production is concentrated close to the event horizon, and is sensitive
to model parameters such as $\dot{m}$.  The pair production rate is also
sensitive to black hole spin $a_*$ and the electron-ion temperature
ratio $\Trat$, but exploring the dependence on these parameters is
beyond the scope of this paper.

We also find that the pair plasma is created with a power-law-like
energy distribution.  Most of the pairs are created in the equatorial plane
of the thick disk because MeV photons created by Compton scattering are
beamed into the equatorial plane.  The pair plasma has negligible effect
on the accretion flow dynamical evolution, consistent with previous
results by \citealt{esin:1999} and \citealt{kusunose:1996}, assuming
that it escapes on the viscous time scale.

Only a few percent of all pairs are created in the magnetized funnel
(black hole magnetosphere), and most of pairs in the funnel are created
near its wall. Pair jets will have spectra with a turnover frequency at around
$\nu_t = 10^{-3} n_{\pm} {\mathcal L}$ Hz (for example, for M87
${\mathcal L}= 4 \times 10^{14}$, and $n_{\pm}=10$, turnover frequency
$\nu_t=10^{12}$ Hz).

We also find that the general relativistic RIAF models are
selfconsistent up to $\dot{m}_{crit} \approx 10^{-6}$, which is
consistent with the $\dot{m}_{crit}=5 \times 10^{-6}$ reported by
\citealt{fragile:2009}.  For higher $\dot{m}$ one must couple the
radiative cooling and forces into the dynamical model.

Models with $\dot{m} < \dot{m}_{crit}$ have force-free, Thomson thin
jets with the Blandford-Znajek luminosity much larger than pair kinetic
luminosity.  In models with very small $\dot{m}$, the pair plasma
density in the funnel is below the Goldreich-Julian density $n_{GJ}$,
suggesting that another process, such as a pair cascades, will operate
and populate the funnel.  

We have applied versions of our model to Sgr A* and to M87.  These
models suggest that $n_\pm > n_{GJ}$ in M87 and $n_\pm < n_{GJ}$ in Sgr
A*, with the important caveat that there are parameters ($a_*$ and
$\Trat$) that we have not varied, and effects (Compton cooling, and
nonthermal electrons) that we have not included.

\bibliographystyle{apj}
\bibliography{local}

\appendix

\section{Synchrotron cooling rates including radiation self-absorption}

The synchrotron cooling rate for a single electron is 
\begin{equation}
\eta^T = \frac{2 e^4  B^2 (\gamma^2-1) \sin^2 \xi}{3 m_e^2  c^3} \label{eq:single_cool}
\end{equation}
where $\xi$ is the pitch angle between electron velocity and magnetic field
(e.g. \citealt{rybicki:1986}). To obtain the total cooling rate from the
thermal population of electrons we integrate Equation~(\ref{eq:single_cool}) against
the relativistic Maxwellian distribution:
\begin{equation}
\frac{dn_e}{d\gamma d \cos \xi} = \frac{n_e \gamma (\gamma^2-1)^{1/2}}{2
  \Theta_e K_2(1/\Theta_e)} \exp(-\frac{\gamma}{\Theta_e})
\end{equation}
The resulting integral over $\cos \xi$ and $\gamma$ is: 
\begin{equation}
\Lambda_S= \frac{4 B^2 e^4 n_e \Theta_e K_3(1/\Theta_e)}{3 c^3 m_e^2 K_2(1/\Theta_e)}
\end{equation}
For $x \ll 1, K_n(x) \rightarrow \Gamma(n)/2(2/x)^n$, so for large $\Theta_e$,
\begin{equation}
\Lambda \rightarrow \frac{16 B^2 e^4 n_e \Theta_e^2}{3 c^3 m_e^2}
\end{equation}
This agree with expression 14 in \citet{wardzinski:2000}. 
For $x\gg1$ ($\Theta_e \ll 1$), $K_n(x) \rightarrow
(\pi/2x)^{1/2} e^{-x}$, and
\begin{equation}
\Lambda_S = \frac{4 B^2 e^4 n_e \Theta_e}{3 c^3 m_e^2}
\end{equation}
The ratio of these two expressions is $4 \Theta_e$; a reasonable 
approximation is
\begin{equation}
\Theta_e \frac{K_3(1/\Theta_e)}{K_2(1/\Theta_e)} \approx (\Theta_e^m 
+ (2\Theta_e)^{2m})^{1/m}
\end{equation}
where $m=4/3$ gives at most $4\%$ error. 

To account for synchrotron selfabsorption, $\Lambda_S$
is multiplied by a factor:
\begin{equation}
f \equiv \frac{1}{\Lambda_{Syn}} \int_0^{\infty} d\nu \int_0^{\pi} \sin \theta
d\theta j_{\nu} \exp(-\tau(\nu,\theta)) \approx \frac{1}{\Lambda_{Syn}}
\int_{\nu_{crit}}^{\infty} d\nu \int_0^{\pi} \sin \theta d\theta j_{\nu} 
\end{equation}
where $j_{\nu}$ is given by Equation~(\ref{emi_leung}), $\Lambda_{Syn}$ is the
first integral without optical depth factor, $\nu_{crit}$ is the frequency
where selfabsorption becomes important. The critical frequency is calculated
numerically from:
\begin{equation}
\kappa_{\nu}(\theta=\pi/2) R =1 \label{nucrit_eq}
\end{equation}
where $\kappa_{\nu}=j_{\nu}/B_{\nu}$, $B_{\nu}$ is the Planck function and
$R=0.1{\mathcal L}$. We find that $f$ well approximated by 
\begin{equation}
f = \frac{1}{2} (\exp(-\frac{X_{crit}}{82}) + \exp(-\frac{X_{crit}}{360}))
\end{equation}
where $X_{crit} = \nu_{crit}/\nu_s$. This fit gives error for $f$ 
less than 1\% up to $X_{crit}=10^2$ and 5\% error at $X_{crit}=10^3$. 

\section{Free-free cooling}

The electron-ion bremsstrahlung cooling rate is \citep{stepney:1983}:
\begin{equation}
 \Lambda_{ei} =n_e n_p \sigma_{T} c \alpha_f m_e c^2 
\left\{ \begin{array}{ll}
\frac{9\Theta_e}{2\pi} (\ln(2\Theta_e \exp(-\gamma_E)+0.42)+1.5 )  & \mbox {$  \Theta_e \ge 1$}; \\
4(\frac{2\Theta_e}{\pi^3})^{0.5} (1+1.78 \Theta_e^{1.34}) &  \mbox {$
\alpha_f^2 \ll \Theta_e<1$}. \end{array} \right. 
\end{equation}
where $\gamma_E=0.5772$ is Euler constant and $\alpha_f$ is the fine
structure constant.  The electron-electron bremsstrahlung cooling rate
is \citep{svenson:1982}
\begin{equation}
\Lambda_{ee} = n_e^2  \sigma_{T} c \alpha_f m_e c^2 
\left\{\begin{array}{ll}
\frac{12}{\pi} \Theta_e (\ln(2\Theta_e \exp(-\gamma_E))+\frac{5}{4}) & \mbox{$  \Theta_e \ge 1$};\\
\frac{5}{6\pi^{1.5}} (44-3\pi^2) \Theta_e^{1.5}
(1+1.1\Theta_e+\Theta_e^2-1.25\Theta_e^{2.5}) & \mbox{ $ \alpha_f^2 \ll \Theta_e < 1$}.\end{array}\right.
\end{equation}
The cooling rates are in units of ${\rm ergs \, s^{-1} \, cm^{-3}}$, and
are consistent within a factor of 2 with those provided by e.g.
\citet{maxon:1972} or \citet{gould:1980}.  Selfabsorption for free-free
emission is negligible.  For $\Theta_e > 1$ the ratio of synchrotron to
bremsstrahlung cooling rate is approximately $\Theta_e^2 /\beta
\alpha_f$.  Synchrotron cooling dominates over the free-free emission in
all of models considered here.

\section{Radiative cooling in MHD code}

Radiative cooling is governed by
\begin{equation}
\frac{du}{dt} = \frac{du}{d\tau}\frac{1}{u^t} = -\frac{\Lambda}{u^t}.
\end{equation}
where $u$ is the internal energy per unit proper volume, $\tau$ is
the proper time, and $u^t$ is the time component of the fluid
four-velocity.  

Numerically $u$ is evolved in an operator-split fashion.  After each
fluid timestep $\Delta t$, $u$ is evolved using the second order scheme
$u_{n+1} = u_n \exp(-\Delta t/\tau_{cool,n+1/2})$ and $\tau_{cool} =
u/\Lambda$.

The cooling rates are calculated in cgs units, and then $\Lambda_{code}
= \Lambda_{cgs} \lunit \tunit^3/\munit$.

\section{Bremsstrahlung emissivity in the radiative transfer calculations}

The emissivity for e-i interactions is \citep{stepney:1983}:
\begin{equation}
  j_{\nu}^{ei} = \frac{dE}{dt dV d\nu d\Omega } = \frac{1}{4\pi} n_i c h \int_{1+\omega}^{\infty} \omega \frac{d\sigma}{d\omega}
  \beta n_e(\gamma) d\gamma \label{brem_emis_ei}
\end{equation}
where $\omega=h\nu/m_ec^2$, $n_e(\gamma)$ is relativistic Maxwellian 
electron energy distribution and the cross-section for this reactions 
is in the ultra-relativistic limit \citep{jauch:1976}. The
$1/4\pi$ factor gives emissivity per unit solid angle.  The integral is
computed numerically using Gauss quadratures. The integration of
Equation~(\ref{brem_emis_ei}) over photon energies and solid angle gives the
total e-i cooling rate, $\Lambda_{ei}$.

The emissivity for e-e emission is also from \citet{stepney:1983}, 
\begin{equation}
 j_{\nu}^{ee}=\frac{1}{4\pi}  n_e^2 \sigma_{T} ch \alpha_f  \Theta_e  
 \exp(-x) G(x,\Theta_e)  \label{emi_ee1}
\end{equation}
where $x= (h \nu/m_ec^2)/\Theta_e$ and $G(x,\Theta_e)$ is given in
\citet{stepney:1983}.
This formula is accurate to $5\%$ over $0.1<\Theta_e<2$.

For $\Theta_e < 0.1$ we use a quadrupole approximation 
(\citealt{maxon:1972}):
\begin{equation}
j_{\nu}^{ee}= \frac{1}{4\pi} \frac{2}{\pi} n_e^2 \sigma_T c h 
\alpha_f  B(x) \sqrt{\frac{2 \Theta_e}{\pi }}  
\exp(-\frac{x}{2}) K_0(\frac{x}{2}) \label{emi_ee2}
\end{equation}
where $B(x)=0.85+1.35\sqrt{x}+0.38x$
and $K_0$ is the modified Bessel function of the second kind. 

For $\Theta_e > 2$ we use the ultra-relativistic approximation
(\citealt{alexanian:1968}, \citealt{maxon:1972}):
\begin{equation}
\begin{array}{ll}
 j_{\nu}^{ee}= \frac{1}{4\pi} \frac{3}{4\pi}  n_e^2 \sigma_{T}  c h
\alpha_f \exp(-x) 
\{\frac{28}{3} + 2 x +\frac{x^2}{2} + 2(\frac{8}{3} 
+ \frac{4}{3} x + x^2) \\
 \times [\lg(\frac{2kT_e}{m_ec^2}) - 0.577]
 -\exp(x) Ei(-x) (\frac{8}{3} -{4}{3}x+x^2) \} \end{array} \label{emi_ee3}
\end{equation}
Formulas~\ref{emi_ee1}, \ref{emi_ee2}, \ref{emi_ee3} connect smoothly at
$\Theta_e$=0.1 and 2.  Bremsstrahlung for e-e interactions dominates
over e-i ones for $\Theta_e > 1$.  Integration of $j_{\nu}^{ee}$ over
frequencies and solid angle gives the total cooling rate,
$\Lambda_{ee}$.

For details of the radiative transfer scheme see \citet{dolence:2009};
we sample the bremsstrahlung radiation field in the same way as for
synchrotron radiation, except that bremsstrahlung is emitted
isotropically in the fluid frame.  For the range of parameters
considered in this work, energy loss by free-free emission is small in
comparison to synchrotron and Compton losses.

\begin{deluxetable}{cccccccc}
\tabletypesize{\scriptsize}
\tablecaption{List of GRMHD models.}
\tablewidth{0pt}
\tablehead{
  \colhead{ID}& \colhead{$a_*$} & \colhead{$m_8$}  &\colhead{$<\dot{m}>_t$}  &
  $L_{Bol}/L_{Edd}$ & radiative & $L_{\pm}/(L_{BZ} \Gamma_{j})$ & note \\
  \colhead{}  &                 &  &\colhead{}  & & efficiency& & 
}
\startdata
    A& 0.94 & $4.5 \times 10^{-2}$ &$2 \times 10^{-9}$&$10^{-11}$&$7 \times 10^{-4}$&$10^{-17}$& 2D\\
    B& 0.94 & $4.5 \times 10^{-2}$ &$6 \times 10^{-9}$&$10^{-10}$&$2 \times 10^{-3}$&$10^{-15}$ & 2D\\
    C& 0.94 & $4.5 \times 10^{-2}$ &$1 \times 10^{-8}$&$4\times10^{-10}$&$4\times10^{-3}$&$10^{-13}$ & 2D\\
    D& 0.94 & $4.5 \times 10^{-2}$ &$5\times 10^{-8}$&$2\times10^{-8}$&0.02&$10^{-10}$ & 2D\\
    E& 0.94 & $4.5 \times 10^{-2}$ &$1\times10^{-7}$&$6\times10^{-8}$&0.04&$10^{-8}$ & 2D\\
\\ 
    F& 0.94 & $4.5 \times 10^{-3}$ & $1 \times 10^{-8}$ & $5\times10^{-10}$&$5\times 10^{-3}$&$10^{-14}$& 2D\\
    G& 0.94 & $4.5 \times 10^{-1}$ & $1 \times 10^{-8}$ & $4\times10^{-10}$&$4\times 10^{-3}$&$10^{-12}$& 2D\\
    H& 0.94 & $4.5 $ & $1 \times 10^{-8}$ & $3\times10^{-8}$&$3\times 10^{-3}$&$10^{-11}$& 2D\\
\\
         &        &                    &                      &\sgra&& \\ 
\\  
    I& 0.94&$4.5\times10^{-2}$&$2.7\times10^{-8}$&$5\times10^{-10}$&$2\times 10^{-3}$&$10^{-11}$&3D-quiescent,$\Trat=3$\\
    J& 0.94&$4.5\times10^{-2}$&$5.3\times10^{-8}$&$1 \times10^{-9}$&$3\times 10^{-3}$&$10^{-9}$&3D-weak flare, $\Trat=3$\\
\\
         &        &                    &                      &M87&&& \\ 
\\  
    K& 0.94 & $30 $  & $ 1.5 \times 10^{-6}$&$3\times10^{-4}$&16.5&0.1&2D - w/o cooling\\
    L& 0.94 & $30 $  & $ 1 \times 10^{-6}$&$3\times10^{-6}$&0.3&$4\times10^{-5}$&2D - w/ cooling\\
\enddata
\tablecomments{\normalsize
From left to right columns are: model ID, dimensionless spin of the black
hole, the black hole mass in units of $M_{\odot}$, the rest mass accretion
rate through the black hole horizon in units of Eddington mass accretion rate
($\dot{M}_{Edd}=2.22 m_8$ ${\rm \MSUN yr^{-1}}$) averaged over later times of
the simulation ($\Delta t = 1500-2000 \tunit$), the Eddington ratio
$L_{Bol}/L_{Edd}$, the model radiative efficiency $\eta=L_{Bol}/\dot{M}c^2$
($L_{Bol}$ is the RIAF luminosity integrated over emitting angles and
frequencies), ratio of Kinetic to electromagnetic luminosity, and comments on
models.  Models I \& J correspond to \sgra\ while K\&L model M87. Run L
accounts for cooling terms in the dynamical solution so the pair
production rate is reduced.\label{tab1}}
\end{deluxetable}

\newpage

\newpage

\newpage
\newpage

\newpage

\newpage

\end{document}